\documentclass[pdflatex, sn-nature]{sn-jnl}

\usepackage{graphicx}%
\usepackage{multirow}%
\usepackage{amsmath,amssymb,amsfonts}%
\usepackage{amsthm}%
\usepackage{mathrsfs}%
\usepackage[title]{appendix}%
\usepackage{xcolor}%
\usepackage{textcomp}%
\usepackage{manyfoot}%
\usepackage{booktabs}%
\usepackage{algorithm}%
\usepackage{algorithmicx}%
\usepackage{algpseudocode}%
\usepackage{listings}%
\usepackage{caption}%
\usepackage{lineno}%

\newcommand{\aap}{Astron. Astrophys.}
\newcommand{\aapr}{Astron. Astrophys. Rev.}
\newcommand{\aaps}{Astron. Astrophys. Suppl. Ser.}
\newcommand{\aj}{Astron. J.}
\newcommand{\apj}{Astrophys. J.}
\newcommand{\apjl}{Astrophys. J. Lett.}
\newcommand{\apjs}{Astrophys. J. Suppl. Ser.}
\newcommand{\araa}{Annu. Rev. Astron. Astrophys.}
\newcommand{\mnras}{Mon. Not. R. Astron. Soc.}
\newcommand{\nat}{Nature}
\newcommand{\natas}{Nat. Astron.}
\newcommand{\pasj}{Publ. Astron. Soc. Jpn.}
\newcommand{\pasp}{Publ. Astron. Soc. Pac.}
\newcommand{\fspas}{Front. Astron. Space Sci.}
\newcommand{\aph}{Astropart. Phys.}
\newcommand{\japa}{J. Astrophys. Astron.}
\newcommand{\raap}{Res. Astron. Astrophys.}

\theoremstyle{thmstyleone}%

\theoremstyle{thmstyletwo}%

\theoremstyle{thmstylethree}%

\begin{document}

\title[Article Title]{Evidence for protostellar jets as a population of hadronic gamma-ray sources}

\author*[1]{\fnm{Javier} \sur{M\'endez-Gallego}}\email{\small jmendez@iaa.es}

\author[1]{\fnm{Rub\'en} \sur{L\'opez-Coto}}\email{\small rlopezcoto@iaa.es}

\author[2]{\fnm{Emma}
\sur{de O\~na Wilhelmi}}\email{\small emma.de.ona.wilhelmi@desy.de}

\author[1]{\fnm{Stefano} \sur{Menchiari}}%\email{smenchiari@iaa.es}
\author[1]{\fnm{Iv\'an} \sur{Agudo}}%\email{iagudo@iaa.es}
\author[1]{\fnm{Rub\'en} \sur{Fedriani}}%\email{fedriani@iaa.es}

\affil[1]{\small \orgdiv{Instituto de Astrof\'isica de Andaluc\'ia (IAA)}, \orgname{CSIC}, \orgaddress{\street{Glorieta de la Astronomía s/n}, \city{Granada}, \postcode{18008}, \country{Spain}}}

\affil[2]{\small \orgdiv{Deutsches Elektronen-Synchrotron}, \orgname{DESY}, \orgaddress{\street{Platanenallee, 6}, \city{Zeuthen}, \postcode{15738}, \country{Germany}}}

%%==================================%%
%% Sample for unstructured abstract %%
%%==================================%%
%\linenumbers

\abstract{ \bf \small
% 200 words
{Stars are born in darkness, deep within cold, dense molecular clouds where gravity drives the collapse of gas and dust, giving rise to protostars, the earliest stages of stellar evolution. 
Once considered purely thermal sources, these young systems are now emerging as sites of energetic non-thermal activity.
While radio synchrotron jets hinted at the presence of relativistic electrons, direct confirmation of proton acceleration remained elusive. 
Here we report a statistically significant detection of $\gamma$ rays from a population of young stellar objects, revealing a Galactic class of Gamma-Loud Protostars.
Observations point towards particle acceleration within protostellar jets, where $\gamma$-ray emission arises from protons interacting with surrounding molecular clouds via pion decay.
We find a correlation between cosmic-ray output and bolometric luminosity, suggesting that particle acceleration scales with the system's mechanical power.
These findings open a new observational window into the role of non-thermal processes in protostellar evolution and suggest that $\gamma$-ray studies of protostars can provide critical insights into accretion, ejection, and feedback in star formation. This previously overlooked emission traces the energetic feedback that young stars inject into their surroundings, shaping the conditions for subsequent star and planet formation.}
}

\keywords{young stellar objects, protostars, gamma rays, cosmic rays, relativistic shocks, protostellar jets}

\maketitle

%%==================================%%
%%             Main Text            %%
%%==================================%%

Stars form within dense molecular clouds through gravitational collapse, giving rise to protostars surrounded by accretion disks \cite{Shu1987, Lada2003, PPVII}. As material spirals inward via accretion disks, angular momentum is redistributed by the disk torques and, in addition, via powerful bipolar jets that emerge along the system's rotational axis. These protostellar jets, observed across optical, infrared, and radio wavelengths, are highly collimated and supersonic, playing a key role in regulating accretion and dispersing excess angular momentum \cite{McKee2007, Reipurth2001}. They are thought to originate from magneto-centrifugal processes operating close to the disk surface \cite{Blandford1982, Pudritz2007, Ferreira2006}, though the precise launching mechanisms —whether from the inner disk or star-disk boundary— remain debated \cite{Shu1994}.

\begin{figure}[h]
     \centering
     \includegraphics[width=0.62\linewidth]{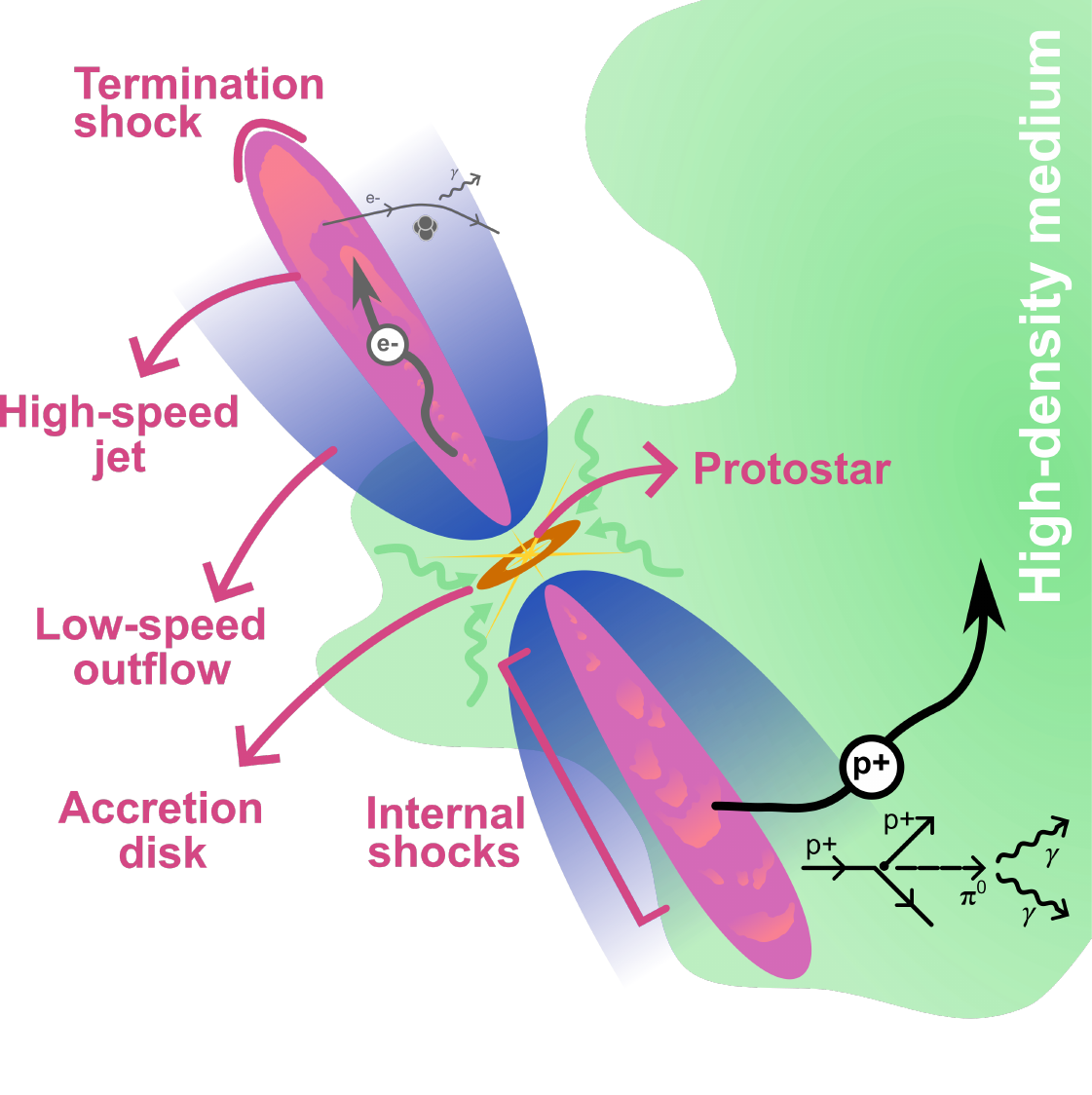}
     \caption{{Illustration of a protostellar jet and its environment}. Particles get accelerated in the shocks formed in the jet or by the accreted material on the protostellar surface. While $\gamma$ rays generated by electrons may be inefficient at high energies, protons can interact with the surrounding medium producing $\gamma$ rays through $\pi^0$ decay. 
     }
     \label{fig:jet_sketch}
\end{figure}

While protostellar jets are primarily studied through their thermal emission from ionized or molecular gas (e.g. \cite{Bally2016, Bally2023, 2023Ray}), an increasing number of systems exhibit signatures of non-thermal radiation, particularly synchrotron emission at radio wavelengths \cite{CarrascoGonzalez2010, Ainsworth2014, 2017Osorio, Kamenetzky2018}. This emission arises from relativistic electrons accelerated in shocks along the jet, indicating the presence of strong magnetic fields and efficient particle acceleration mechanisms \cite{RodriguezKamenetzky2025} akin to those operating in more prominent astrophysical jets, such as blazars or microquasars. Linear polarization measurements and spectral index mapping confirm the synchrotron origin in several cases, offering a powerful probe of jet magnetic topology and energy dissipation. Non-thermal emission from protostellar jets can alter disk parameters by impacting the chemistry, accretion dynamics, and dust properties \cite{2007A&A...468..515G}, thereby influencing star formation and potentially even planet formation \cite{Tychoniec_2018}. The detection of non-thermal emission challenges classical models of stellar evolution and opens a new observational window onto the interplay between magnetic fields, turbulence, and shocks during the earliest stages of star formation. Moreover, it suggests that young stellar objects (YSOs) can act as cosmic accelerators, with implications for the origin of cosmic rays (CRs) in star-forming regions. Systematic surveys are increasing the sample of non-thermal jet candidates \cite{Obonyo2024}, yet only a handful of young stellar objects have been unambiguously confirmed as synchrotron emitters where relativistic electrons are likely accelerated in the jet. Even rarer are systems detected at $\gamma$-ray energies, which penetrate the high column densities of the natal cloud and directly trace ultrarelativistic particles (e.g. \cite{Araya2022, deOnaWilhelmi2023, MendezGallego2025}). From a theoretical perspective, protostellar jets and magnetospheric accretion shocks can provide natural environments for particle acceleration via diffusive shock acceleration or magnetic reconnection \cite{Padovani2009, BoschRamon2010, Araudo2021}. These models predict spectral energy distributions peaking close to the GeV range, with flux levels that may be potentially detectable by current and next-generation instruments, such as CTAO \cite{CTAO2019}. However, up to now, the observational $\gamma$-ray detections could only identify protostellar jets as potential $\gamma$-ray emitters, without firmly associating YSOs themselves with $\gamma$-ray sources. Moreover, current $\gamma$-ray analyses do not constrain the particle population producing them, having as possible origin ultrarrelativistic electrons emitting $\gamma$ rays through relativistic Bremsstrahlung or -even more relevant to the elusive origin of the Galactic CR population- accelerated protons that emit $\gamma$ rays through pion decay after interacting with the interstellar material present in these regions (see Fig. \ref{fig:jet_sketch}). 

% Add Figure 1

\begin{figure}[h]
    \centering
    \captionof{table}{Association parameters. Best-fit priors and resulting association statistics derived from the likelihood ratio method.}
    \includegraphics[width=0.3\linewidth]{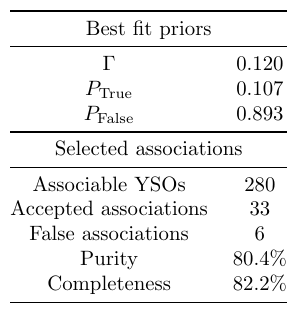}
    \label{tab:assoc_params}
\end{figure}

\begin{figure}[h!]
    \centering
    \captionof{table}{{Bayesian association of GLPs and their corresponding parameters}. Previous $\gamma$-ray associations recovered from literature are indicated with ($\ast$). Extra columns regarding \textit{Fermi}-LAT flux points, non-thermal particle energy, and particle distribution parameters can be found in the digitalized version of the table.}
    \includegraphics[width=0.97\linewidth]{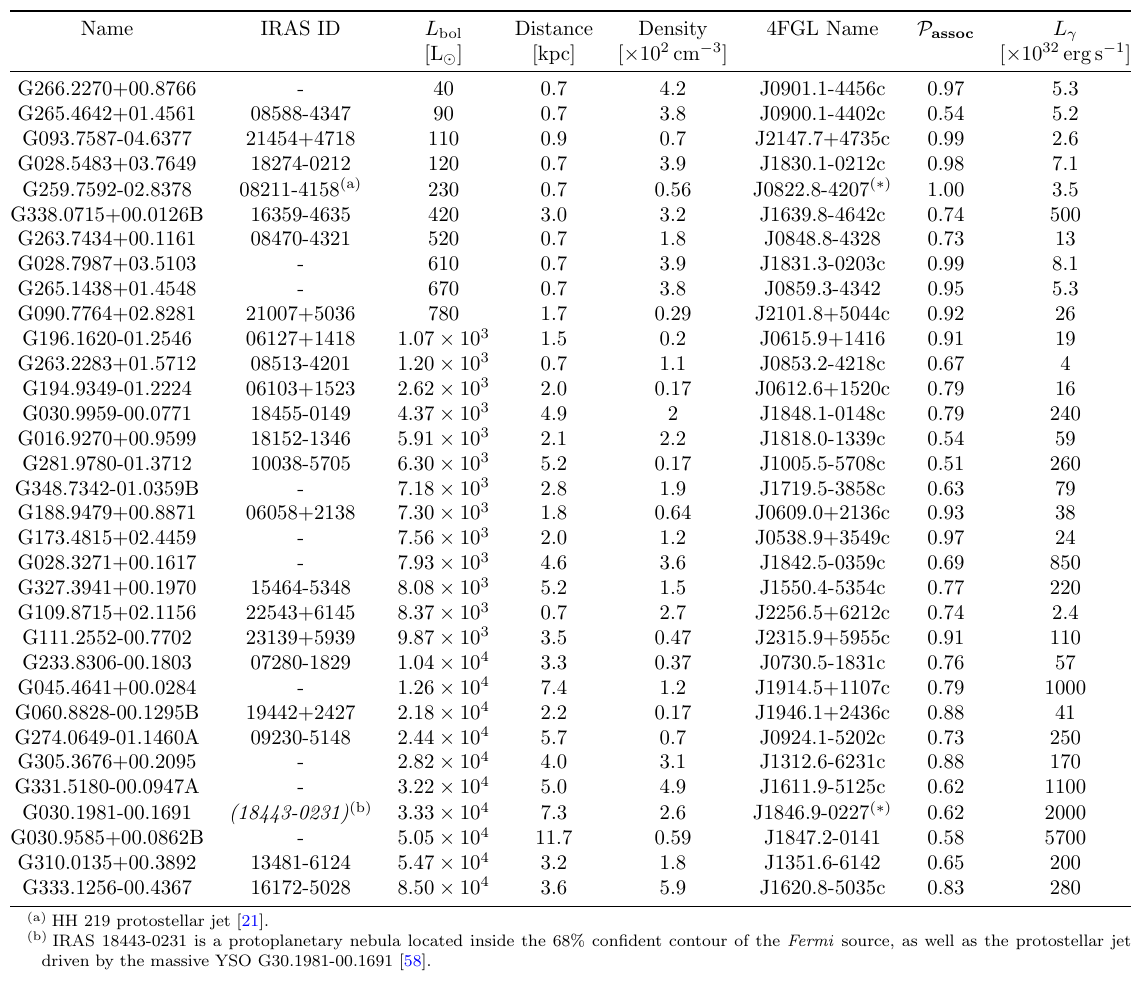}
    \label{tab:GLPs}
\end{figure}

In this work, we used public data from the \textit{Fermi}-LAT 4FGL data release (DR)~4 catalog of point-like $\gamma$-ray sources. 
The 4FGL DR4 catalog \cite{Abdollahi2020, Ballet2023} is the most recent and complete survey of $\gamma$-ray sources, containing all-sky results from 14-year data taking of \textit{Fermi}-LAT. 
We compare the distribution between these $\gamma$-ray sources and the positions of YSOs from the Red MSX (\textit{Midcourse Space Experiment}) Source (RMS) survey \cite{2013Lumsden}. The method used to carry out this association is explained in detail in 
\hyperref[sect:LR_assoc]{Methods}, where we optimized the fraction of associated sources among the initial sample using a Bayesian approach based on likelihood comparison. Since we did not perform any morphological analysis, and we are only interested in point-like $\gamma$-ray sources, the standard three-dimensional analysis of \textit{Fermi}-LAT data (which consists in modelling a region of interest that contains angular coordinates plus energetic dimension of sources and spectra) retrieves a positional uncertainty of the order of $\sim$0.2$^\circ$.
As a result, we statistically associate and identify 33 pairs of YSOs and $\gamma$-ray objects with a purity of $\sim$80\% and a given probability of association ($\mathcal{P}_{\rm assoc}$) for each case (Table \ref{tab:assoc_params}). The amount of source associations deviates approximately thirteen standard deviations ($\sigma$) from what is expected from simulations of randomly distributed sources. A list of Gamma-Loud Protostars (GLPs) selected by this methodology and their properties can be found in Table \ref{tab:GLPs}. 
We validated these results by performing an updated study on the positional cross-matching between YSOs and $\gamma$-ray sources compiled in the 4FGL DR4 catalog based on \cite{MunarAdrover2011} (Extended Table~\textcolor{blue}{1}), and using a \emph{longitude resonance} approach \cite{2009Hillas} (see Extended Fig.~\textcolor{blue}{1}
%\ref{fig:hillas_shift}
and \hyperref[sect:validation]{Validation} section).

\begin{figure}[h!]
     \centering
     \includegraphics[width=0.85\textwidth]{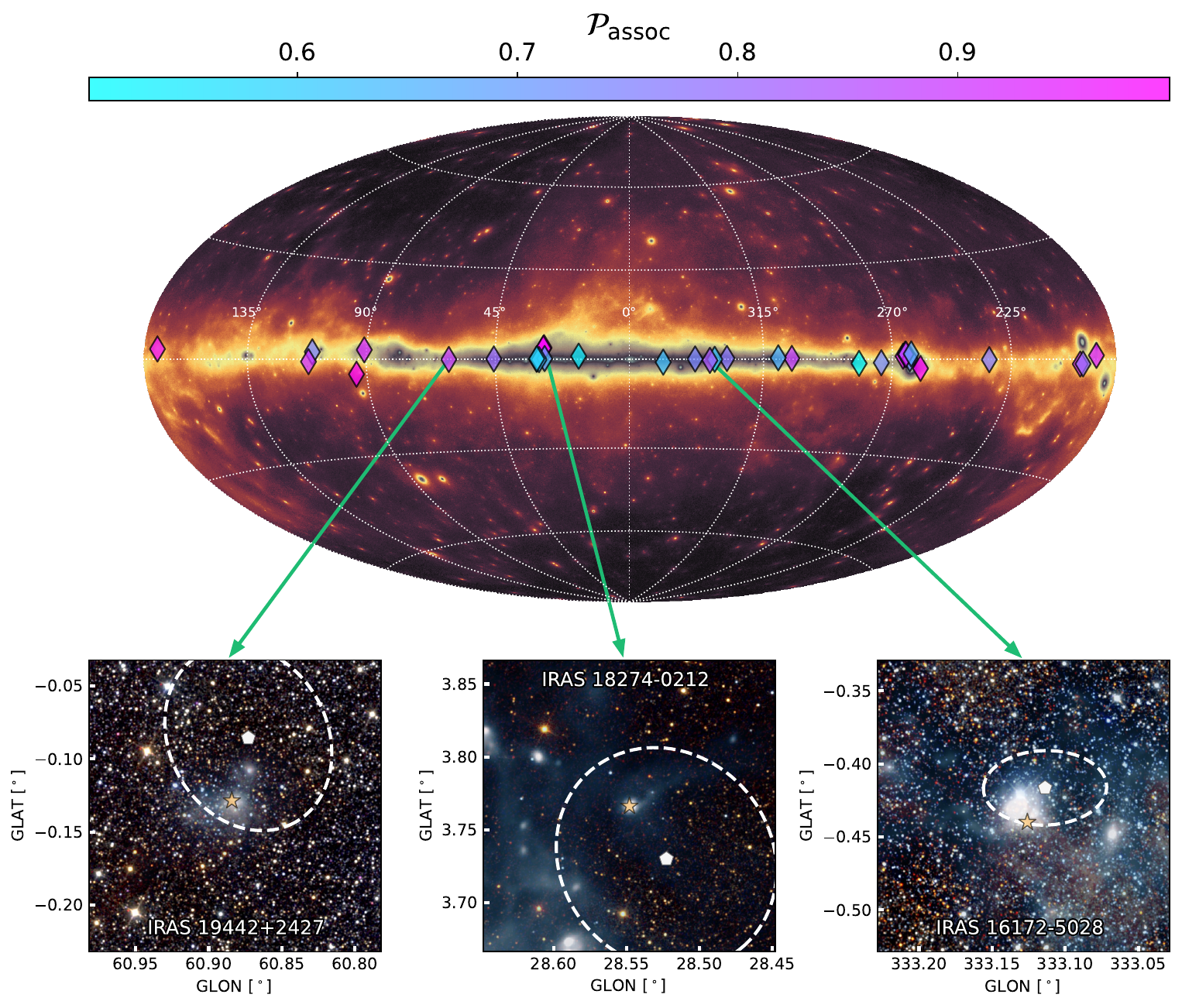}
     \caption{{Location of the 33 GLPs in the 4FGL skymap}. Top panel displays the all-sky map from Fermi-LAT above 100 MeV, where diamonds indicate the position of the associated GLPs in this work. The color code reports the probability of association of each YSO (yellow star) and its $\gamma$-ray pair. Highlighted panels show three examples of GLPs identified in this work. Dashed contours represent the 95\% confidence area of the $\gamma$-ray source position (white pentagon). The background image in the lower panels is a combination of \textit{Two Micron All Sky Survey} (colored) image with \textit{Herschel} observations at 160 $\mu$m (blue shades).}
     \label{fig:skymap}
 \end{figure}

We also performed a multiwavelength study to search for possible association from pulsars, pulsar wind nebulae, supernova remnants, young stellar clusters and extragalactic sources and found that only three of the 33 GLPs have another possible counterpart within the 99\% confidence area from the location of the 4FGL source, ruling out other source types as the origin of this $\gamma$-ray emission.
Figure \ref{fig:skymap} shows the distribution of the detected sample of what we are designating GLPs along the Galaxy \cite{2013Atwood, 2006Skrutskie, 2010Poglitsch}.

% Add Figure 2

The identification of such a small sample of GLPs, which constitute only a fraction of $\sim$11\% among all the possible associations between the RMS survey and the 4FGL catalog, may be the result of two different effects. First of all, the low $\gamma$-ray fluxes these sources exhibit render it difficult to detect objects located at large distances (see Supplementary Sect. \ref{sect:bias}). Secondly, the non-detection of YSOs with similar characteristics and at similar distances to GLPs points towards the possibility that only a selected portion of YSOs is in an active state where $\gamma$-ray detection is possible. YSOs have been demonstrated as dynamical sources where violent accretion episodes are frequently generating outbursts and flares that can cause significant shocks in the material falling onto the protostar or increase the jet activity, resulting in new shocks as the ejecta propagates at high speed along the collimated outflow \cite{2017Hunter, 2018Cesaroni, 2023Fedriani}. As a consequence, particles may not be continuously accelerated, but rather CRs may be produced in concrete periods during the protostellar lifetime, being detected only those protostars whose accretion rate has been relatively high and stable.

YSOs undergoing active accretion are known to produce collimated protostellar jet \cite{Pudritz2007, 2024Bally}. Consistently, we found that 21 of the 33 sources compiled in Table \ref{tab:GLPs} exhibit jet-like signatures (see Supplementary Sect. \ref{sect:assoc+}). For the remaining objects, the absence of confirmation appears to result from the lack of targeted observations rather than an actual non-detection of jet structures. In these cases, future dedicated observations using molecular lines at radio and sub-millimetre wavelengths would be valuable for definite identify the acceleration origin of the detected CR excess.

In most of the identified GLPs, the detection of $\gamma$ rays reaches energies of 10 GeV or higher. The association of GeV $\gamma$-ray emission with protostellar sources implies the presence of relativistic particles reaching tens to hundreds of GeV, offering critical insights into both the nature of the particles and their acceleration site. Surface shocks near the protostar are constrained in their capacity to accelerate particles: electrons are limited to energies of a few hundred MeV, and protons to $\sim$30 GeV, under typical conditions \cite{Padovani2015}. These limits, consistent with independent estimates \cite{2018Gaches}, fall short of accounting for the observed $\gamma$-ray flux. In addition, the maximum energies reached in these systems can be attained by the highly magnetized ($B_{\rm 200\,km/s}\sim$0.1 mG for $E_{\rm max}$=10 GeV) and fast velocities that are present in some protostellar jets, presenting a more favourable environment for acceleration (as it is explained in detail in Sect. \ref{sect:kinetics} of Supplementary Information). These magnetized jets also support models of jet formation via magneto-centrifugal mechanisms (see \cite{Blandford1982, 1986Pudritz} for further details). Consequently, since both theoretical and observational frameworks propose that high-energy particles are accelerated in jet shocks (see \cite{Padovani2015} and \cite{CarrascoGonzalez2010}), we conclude that our sample of GLPs indicates the presence of strong protostellar jets originating in young accreting systems.

Other important topic covered by this work is the nature of the particles producing the observed emission. Synchrotron losses (the cooling time for electrons with energies of 100~GeV at $\sim$mG magnetic fields is as short as $\sim$125~yrs) restrict electron maximum energies even in jets \cite{Padovani2015}, contrary to protons, that can reach substantially higher energies. Under perpendicular shock geometries and at larger distances from the protostar, proton energies may extend to $\sim$100 GeV or beyond \cite{Padovani2016}, consistent with other models predicting acceleration up to hundreds of GeV \cite{Araudo2021}, TeV \cite{BoschRamon2010}, or even higher scales \cite{Araudo2007}. These energies are the compatible with the emission observed in our population of sources and render hadronic interactions as the most plausible emission mechanism. 
In the same way, the non-thermal kinetic energies inferred from the particle distributions reproducing the observed $\gamma$-ray spectra favor a hadronic scenario, which requires acceleration efficiencies of order $\sim$10\%, whereas an electron-dominated scenario would demand significantly higher efficiencies than the canonical value of $\sim$1\% (see Supplementary Sect. \ref{sect:kinetics} for further details).

\begin{figure}[h]
     \centering
     \includegraphics[width=0.7\linewidth]{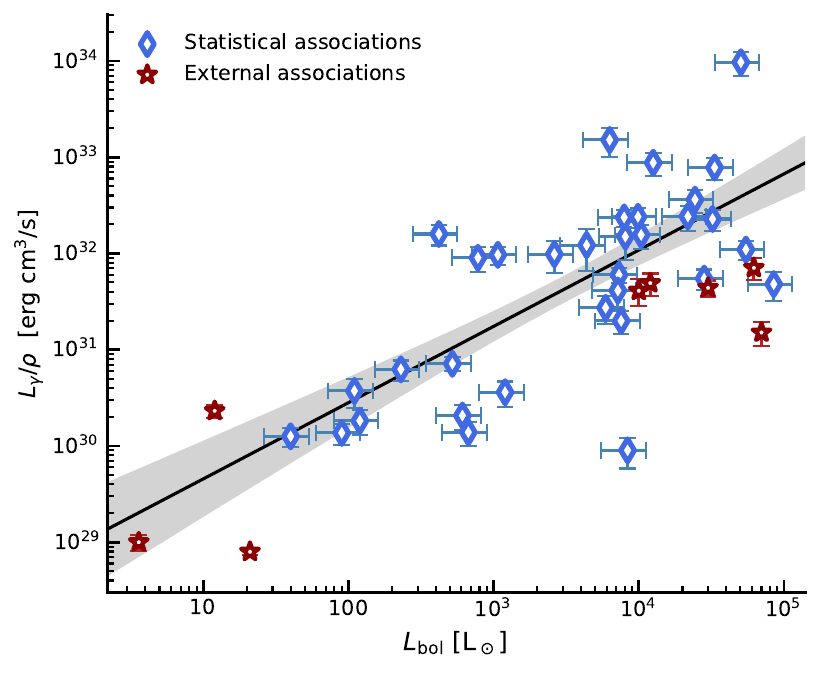}
     \caption{{Bolometric luminosity versus $\gamma$-ray luminosity per unit of density}. Blue diamonds indicate the 33 GLPs identified in this study. Red stars represent externally associated sources, suggesting that the correlation could extend to other protostellar jets. The black line shows the best fit to the blue diamonds, with the shaded area representing its uncertainty. Error bars account for a relative uncertainty of 34\% in the RMS bolometric luminosity \cite{2011Mottram}. For the y-axis, only the uncertainties in the measured $\gamma$-ray fluxes are considered.}
     \label{fig:Lgamma_vs_Lbol}
 \end{figure}

Since thermal radiation from young protostars is mainly driven by accretion and gravitational collapse of the interstellar material to the dense cores where stars are born, we consider the bolometric luminosity of the protostellar system ($L_{\rm bol}$) as a proxy for the protostellar jet kinetic power \cite{Cabrit2007}. A confirmation of this proportionality can be found in works correlating $L_{\rm bol}$ to quantities like the momentum rate of the jet (Figs. 8 and 9 of \cite{Anglada2018}) or H$_2$ (Fig. 9 of \cite{Caratti2015}), both related with the jet mechanical activity.  
Therefore, Fig. \ref{fig:Lgamma_vs_Lbol} presents a phenomenological relation of the capability of YSOs for producing relativistic particles versus the $L_{\rm bol}$ of the systems as an indication of their jet mechanical power, whose exponential index is consistent with the relation of $L_{\rm bol}$ versus jet mechanical power (see Supplementary Sect. \ref{sect:jet_power}).

% Add Figure 3

\begin{figure}[h]
    \centering
    \captionof{table}{{Parameters of sources associated in the literature}. References are indicated in the `Name' column.}
    \includegraphics[width=0.95\linewidth]{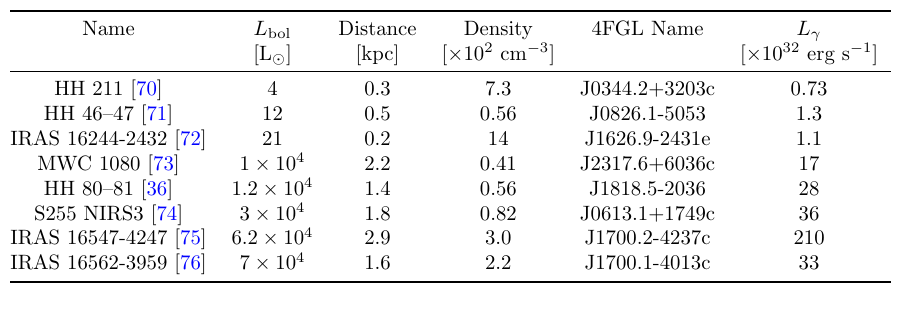}
    \label{tab:literature_associations}
\end{figure}

The $\gamma$-ray luminosity ($L_\gamma$; integrated at higher energies than 100~MeV) produced inside YSO environments via either Bremsstrahlung or pion decay will be directly proportional to the amount of accelerated particles and the ambient density of the medium where these relativistic particles are interacting ($\rho$). Hence, with $L_{\rm \gamma}/\rho$ we are characterizing the CR abundance in these sources and comparing it with the power of the whole system. 
The Pearson's coefficient for our entire sample of selected GLPs is $\sim$0.7.
It is important to note that the observed $\gamma$-ray emission does not present any correlation with our calculations on the ambient density in these regions (see Extended Fig.~\textcolor{blue}{2}), ruling out that the bulk of the emission comes from Galactic diffuse CRs illuminating molecular clouds. The clear correlation proves that there is a sample of YSOs within the Milky Way accelerating CRs during its formation stage. The correlation is also followed by the externally associated sources of Table \ref{tab:literature_associations} (see \hyperref[sect:external_assoc]{External associations} section). Additionally, assuming that all our associated sources have jets, the relation derived from Fig. \ref{fig:Lgamma_vs_Lbol} indicates a strong dependence on the CR production and the kinetic energy of the protostellar jet, as predicted by DSA acceleration models. In these models, part of the kinetic energy contained in shocks is transformed into CRs, with typical maximum efficiencies of the order of 10\% \cite{Araudo2021, 2024Peron}. The $\gamma$-ray luminosity of these objects is comparable to that observed in typical Galactic CR accelerators such as evolved, interacting SNRs \cite{2016ApJS..224....8A} and also young stellar clusters \cite{Aharonian2019}.

Equation (\ref{eq:Lgamma_vs_Lbol}) represents the relation derived from Fig. \ref{fig:Lgamma_vs_Lbol}. 
The dependency of our findings on $L_{\rm bol}$ is compatible with the independent result obtained in radio wavelengths when comparing $L_{\rm bol}$ versus the jet mechanical power \cite{2018AtlasGal}, concluding that the selected YSO can be $\gamma$-ray emitters. 
Additionally, $\rho$ plays a remarkable role in characterizing the $\gamma$-ray production. With typical densities of $\sim$100 cm$^{-3}$ and assuming a directly proportional relation between $L_{\rm bol}$ and $L_\gamma$ --where the exponential index is fixed to one--, we obtain that the $L_{\gamma}$--$L_{\rm bol}$ ratio is of the order of $\sim$10$^{-4}$.
In addition, literature indicates that the lifetime scales for a protostellar jet typically last from $10^4$ to $10^6$ yrs, and the kinetic power of the jet usually tends to be of the order of $\sim$10\% of $L_{\rm bol}$ or above \cite{Podio2021}.
Therefore, the total kinetic energy contained in the particle distribution (see Supplementary Sect. \ref{sect:kinetics}) is easily achievable by assuming an acceleration efficiency of 10\% (that indicates the fraction of energy that is transferred from shock to CRs), which is in good agreement with theoretical frameworks.

\begin{equation}
    \frac{L_{\gamma}}{\rm erg \, s^{-1}} = 10 ^{28.9\pm 0.5} \left( \frac{\rho}{\rm cm^{-3}} \right) \left( \frac{L_{\rm bol}}{\rm L_\odot} \right) ^{0.79\pm 0.15}
    \label{eq:Lgamma_vs_Lbol}
\end{equation}

The discovery of GLPs marks a pivotal advance in our understanding of the non-thermal high-energy processes that accompany stellar birth. By statistically associating $\gamma$-ray sources with YSOs, we provide compelling evidence that protostars can act as efficient CR accelerators, with particle energies reaching hundreds of GeV and beyond. Considering the assumptions and estimations of this work, the observed correlation between the $\gamma$-ray and bolometric luminosity underscores a direct link between accretion-driven jet mechanical energy and non-thermal particle production. These findings not only challenge the long-standing view of protostars as purely thermal emitters but also position them as active agents in shaping their environments through energetic feedback \cite{Bally2016}. As $\gamma$-ray observatories continue to refine their sensitivity and resolution, the emerging population of GLPs offers an additional window into the interplay between accretion, outflows, and magnetic fields in the early stages of stellar evolution. In this light, non-thermal processes are not peripheral phenomena anymore; they are integral to the star formation narrative, and $\gamma$ rays arise as a unique window for disentangling the cosmic impact that this population of particle accelerators may have. 
As we lift the veil on the high-energy depths of protostars, a paradigm emerges: non-thermal processes could lie at the heart of some star formation stages.

%%==================================%%
%%             Methods              %%
%%==================================%%

\vspace{5mm}

\section*{Methods}
\label{sect:methods}

\section{Source association}
\label{sect:LR_assoc}
The conclusions of this work are based on the association of $\gamma$-ray sources detected by the \textit{Fermi} satellite and YSOs compiled in catalogs provided infrared observations, where star-forming cores exhibit the peak of their thermal emission. Accordingly, we based our study on the 4FGL catalog of point-like $\gamma$-ray sources compiled from photons detected by \textit{Fermi}-LAT in the energy range between $\rm 100 \, MeV$ and $\rm 100 \, GeV$. The containment angle for event reconstruction of \textit{Fermi}-LAT is strongly energy dependent, with typical values being $\sim$5$^\circ$ above 100 MeV, $\sim$0.8$^\circ$ above 1 GeV and $\sim$0.1$^\circ$ above 20 GeV, retrieving source positional uncertainties of the order of 0.2$^\circ$. Additionally, we mask all previously associated sources in this catalog, thereby removing all the $\gamma$-ray sources linked to other known $\gamma$-ray emitters (e.g. active galactic nuclei, pulsar wind nebulae, supernova remnants, pulsars...), leaving only the unassociated sources and those mark as `unknown' (see Supplementary Table~\ref{tab:catalogs}).
In the same way, we use the positions of the sources classified as YSOs and HII/YSOs in the RMS Survey of the Galactic young stellar population \cite{2013Lumsden}, excluding all the sources with $L_{\rm bol} = 0$. Given that the average positional uncertainties for the $\gamma$-ray sources are of the order of arcminutes and protostellar jets typically fall within this angular range, we treat our YSO counterparts as point-like sources. Thus, YSO positional uncertainties can be neglected within the relatively large positional uncertainty regions of the $\gamma$-ray sources. Both survey coverages are shown in Supplementary Sect. \ref{sect:assoc+}.

Each $\gamma$-ray source is initially matched with a counterpart defined as the closest YSO at a certain angular distance ($\hat{r}$). Similarly, each $\gamma$-ray source has a positional uncertainty ($\sigma$) depending on the projected angle ($\phi$) described by its linked counterpart, defined as follows:

$$ \sigma = \frac{\sigma_{\rm 95, \, a} \times \sigma_{\rm 95, \, b}}{\sqrt{\left( \sigma_{\rm 95, \, a} \sin \phi \right) ^2 + \left( \sigma_{\rm 95, \, b} \cos \phi \right) ^2}} \times \frac{1}{\sqrt{-2 \ln 0.05}}, $$
where $\sigma_{\rm 95, \, a}$ and $\sigma_{\rm 95, \, b}$ represent the semimajor and semiminor axis of the 95\% positional confidence contour for each $\gamma$-ray source collected in the 4FGL DR4 catalog, and the right hand side term is used for converting the 95\% confidence contour into the standard deviation.

Consequently, following the likelihood ratio methodology for source association described in \cite{1977deRuiter, 1992Sutherland}, we study the statistical distributions of $\hat{r}$ and $\sigma$ for determining whether our $\gamma$-ray sources are associated with the corresponding YSOs (see Supplementary Fig.~\ref{fig:assoc_scheme}\textcolor{blue}{a}).

In this context, a YSO counterpart may be located near a $\gamma$-ray source either by chance or due to a true physical association. Sources that are close by chance follow the next marginal distribution:

$$ \mathcal{P}_{\rm Random} (\hat{r} | \rho_{\rm cpt}) = 2\pi \rho_{\rm cpt} \hat{r} \cdot \exp \left( -\pi \rho_{\rm cpt} \hat{r}^2 \right), $$
which results from a Poisson distribution describing the probability of finding no counterpart up to an angular distance $\hat{r}$, for a given surface density of counterparts ($\rho_{\rm cpt}$). $\rho_{\rm cpt}$ is derived from the multiplication of the normalized positional distributions of the YSOs in the RMS survey (see Supplementary Fig.~\ref{fig:cpt_distrib}), taking into account the total number of YSOs that we have in our sample. Therefore, the counterpart density can be calculated locally depending on the region that we are analyzing.

Alternatively, the probability of finding a physically associated YSO at a distance $\hat{r}$ from a $\gamma$-ray source is characterized by the positional uncertainty, which follows a Rayleigh distribution:

$$ \mathcal{P}_{\rm True} (\hat{r} | \sigma ) = \frac{\hat{r}}{\sigma^2} \cdot \exp \left( - \frac{\hat{r}^2}{2 \sigma^2} \right).$$

Above, we defined two likelihoods to describe the two different kind of cases that are possible to find when looking for associated sources: either the $\gamma$-ray source is physically associated to the YSO counterpart ($\mathcal{P}_{\rm True}(\hat{r}_i | \sigma_i )$) or the pair of counterpart and $\gamma$-ray source is associated by chance ($\mathcal{P}_{\rm Random}(\hat{r}_i | \rho_{\rm cpt_i} )$). Both distributions will be present in our sample (Supplementary Fig.~\ref{fig:assoc_scheme}\textcolor{blue}{b}) and they will vary on each different pair, since all the parameters included will change for each case.

Using the Bayes theorem, we construct a posterior ($\mathcal{P}_{\rm Assoc}$) that will return the probability for each $\gamma$-ray source to be truly associated with its corresponding counterpart. To do this, we have to introduce two priors representing the probability of having a true physical association ($P_{\rm True}$) and a false initial match by chance ($P_{\rm False}$). Consequently, both cases are supplementary and $P_{\rm True}+P_{\rm False}=1$. Then, the probability of having a physical association is given by:

$$\mathcal{P}_{\rm Assoc} (\hat{r} , \sigma , \rho_{\rm cpt}) = \frac{\mathcal{P}_{\rm True} (\hat{r} | \sigma ) P_{\rm True}}{\mathcal{P}_{\rm True} (\hat{r} | \sigma ) P_{\rm True} + \mathcal{P}_{\rm Random} (\hat{r} | \rho_{\rm cpt}) P_{\rm False}}.$$

Defining the likelihood ratio as $\mathcal{L}=\mathcal{P}_{\rm True} / \mathcal{P}_{\rm Random}$ and the prior ratio as $\Gamma = P_{\rm True} / P_{\rm False}$, we rewrite the last equation to obtain the simplified Eq. (\ref{eq:P_assoc}) for $\mathcal{P}_{\rm Assoc}$, where $\Gamma$ remains as the only free parameter that must be fit to our data.

\begin{equation}
    \mathcal{P}_{\rm Assoc} (\Gamma | \hat{r} , \sigma , \rho_{\rm cpt}) = \frac{1}{1+ 1 / \left( \Gamma \mathcal{L}(\hat{r} , \sigma , \rho_{\rm cpt})\right)}.
    \label{eq:P_assoc}
\end{equation}

Supplementary Fig.\ref{fig:cpt_distrib} shows a clear $\rho_{\rm cpt}$ dependency on the Galactic latitude in our YSO sample, as most of the sources will be located in the Galactic plane, finding no source above $|b| > 6^\circ$. In addition, we also observe a dependency on the Galactic longitude, since the density of YSOs grows as we approach the center of the Galaxy. The catalog of YSOs does not include sources in the inner part of the Galaxy ($|\ell| < 7^\circ$), thus we mask the same region on the LAT catalog (see Supplementary Table~\ref{tab:catalogs}). In a minor order, we also detect the Galactic spiral arms producing some peaks in the marginal distribution of the Galactic longitude. The observed distribution of YSO is used to derive the parameter $\rho_{\rm cpt}$, which depends on the location in the Galactic plane.

To guarantee a correct number of associated sources to develop a correct statistical analysis, we assume that a pair is associated when $\mathcal{P}_{\rm Assoc} >0.5$. Note that for high values of $\rho_{\rm cpt}$, the likelihood of finding a random source within the $1\sigma$ area increases significantly, making some cases indistinguishable from true associations and therefore effectively unidentifiable, so we masked them. The resulting number of associated sources among our sample of counterparts will depend on the prior probabilities (represented in Eq. (\ref{eq:P_assoc}) by $\Gamma$).

To optimize the prior values for our specific case, we compute the expected number of false associations among the accepted associations as a function of $\Gamma$. We compare our results with the accepted associations that our methodology would obtain from a randomly distributed counterpart population, preserving the original density distribution along the Galaxy. Supplementary Fig.~\ref{fig:Gamma} illustrates the good agreement between the number of false associations estimated using both a Poisson distribution and a Monte Carlo sampling of 10,000 mock counterpart populations (each population containing 827 mock positions, the same amount that we found in Supplementary Table~\ref{tab:catalogs}). 
The cross point where the false associations among the accepted sample match the number of mock associations is adopted as the correct value for $\Gamma$ and, thus, for the priors.

Based on the results shown in Supplementary Fig.~\ref{fig:Gamma}, Table \ref{tab:assoc_params} presents the prior parameters that best fit our sample of linked 4FGL sources and YSOs. The best-fit priors indicate that approximately 11\% of the total YSO population is associated. From the initial YSO sample, only about 34\% are considered for association. The remainder are excluded due to the high local density of counterparts in their regions. Finally, we identify 33 accepted associations, including  6 false associations. For each $\gamma$-ray source-YSO pair, we obtain a distinct value of $\mathcal{P}_{\rm Assoc}$, depending on the specific conditions of each case.

The synthetic populations produced for computing the number of false associations expected at each $\Gamma$ indicate that the number of associated sources at the optimal prior ratio value is significantly higher than the predicted value of false associations, with $\sim$13$\sigma$ of significance.

Following the described methodology, we accepted associations with $\mathcal{P}_{\rm Assoc} > 0.5$ despite the standard value for this type of association being around 0.8. In this way, our sample of selected GLPs will have lower reliability (more false associations among the selected sample), but higher completeness. In total, our association methodology estimates that approximately 40 YSOs from the initial RMS catalog are associated with $\gamma$-ray sources. However, YSOs not included in the RMS catalog are not considered in these results. In the same way, we are interested in faint $\gamma$-ray sources whose position may be shifted further than the $2 \sigma$ positional uncertainty due to systematics, as in the case of HH 80-81 \cite{MendezGallego2025}. While these challenges can be addressed through dedicated analyses of individual sources, the complexity of such an approach compels us to rely on the standard 4FGL catalog.

In the next subsections, we validate this methodology and discuss about other potential candidates to become GLPs.

\subsection{Validation}
\label{sect:validation}

The association of YSOs with gamma-ray sources was previously attempted by other authors \cite{MunarAdrover2011}. Using the Fermi First Year Catalog \cite{2010Abdo}, 12 massive YSOs were positionally coincident, with 4 false associations among them and a confidence level of $\sim$4$\sigma$.

The 4FGL-DR4 catalog has 5 times more sources detected along the 14 years of data taking and many improvements on the $\gamma$-ray event reconstruction after the Pass 8 Release 3 data processing \cite{2018Bruel}. As a consequence, the current status of the newest \textit{Fermi}-LAT 4FGL catalog has significantly more sources with a better sensitivity and a more accurate direction reconstruction than the Fermi First Year Catalog.
To validate our results, we reproduce the methodology of \cite{MunarAdrover2011} to cross-check the potential existence of a population of YSOs emitting in $\gamma$-rays based on the updated \textit{Fermi} 4FGL catalog.

Starting from the number of not masked sources specified in Supplementary Table~\ref{tab:catalogs}, we cross-match the RMS survey with the 4FGL catalog to obtain 315 pairs of YSOs and $\gamma$-ray sources. We only consider the closest YSO to each $\gamma$-ray source and vice versa, reproducing a similar case as in the likelihood ratio method. 

As explained in Supplementary Fig.~\ref{fig:assoc_scheme}, we consider elliptical errors in the $\gamma$-ray source position, with a certain degree of inclination. Therefore, we rotate the system of reference to define the coordinates of our YSO counterpart for obtaining the distance with respect to the semimajor and semiminor axes of the uncertainty ellipse:

\begin{equation*}
    \begin{cases}
        \hat{r}_{\rm a} = \left( \Delta \alpha \right) \cos (\delta ) \sin (\theta ) - \left( \Delta \delta \right) \cos (\theta ) \\
        \hat{r}_{\rm b} = \left( \Delta \alpha \right) \cos (\delta ) \cos (\theta ) + \left( \Delta \delta \right) \sin (\theta ),
    \end{cases}
\end{equation*}
where $\Delta \alpha$ ($\Delta \delta$) is the separation in RA (Dec), and $\theta$ is the position angle of the ellipse. Using this projection and the equation of an ellipse, one can check whether the counterpart is situated inside the elliptical errors of the closest $\gamma$-ray source, building a statistic as follows:

$$ S = \sqrt{\left( \frac{\hat{r}_{\rm a}}{\sigma _{\rm a}} \right)^2 + \left( \frac{\hat{r}_{\rm b}}{\sigma _{\rm b}} \right)^2  }, $$
where $\sigma_{\rm a}$ and $\sigma_{\rm b}$ represent the semimajor and semiminor confidence contours of the acceptance area surrounding the $\gamma$-ray source for claiming a source positional coincidence. Therefore, we can use different acceptance areas considering the several available sizes of confidence contours.

If $S<1$, we accept the positional coincidence, since the respective YSO is falling inside the defined ellipse. In the same way, we test the significance of this amount of positional coincidences producing 10,000 Monte Carlo populations from the Galactic distributions of Supplementary Fig.~\ref{fig:cpt_distrib}, again producing 827 mock YSOs per each Monte Carlo sampling.

Results of Extended Table~\ref{tab:positional_coincidences} 
strengthen the findings regarding the GLP population found through the likelihood ratio methodology of \hyperref[sect:LR_assoc]{Methods}. Under the standard acceptance area formed by the 95\% confidence contour, we recover 32 coincidences out of the 33 associations selected in the likelihood comparison methodology. The source excluded is the furthest source to the Galactic plane, where the density of counterparts is so low that the likelihood ratio methodology obtains more favourable probabilities of association.

We further tested the correlation by using the approach described in \cite{2009Hillas}. In such an approach, the position of the sources along the longitude, keeping the latitude fixed, is shifted, to test  whether the correlation is sensitive to observational bias or non-uniform longitude distribution — what the author calls a right ascension (longitude) \emph{resonance}. A resonance in longitude alignment indicates a genuine correlation. 
The result is shown in Extended Fig.~\textcolor{blue}{1}. A significant minimum close to $\Delta \ell = 0^\circ$ when compared with the self-noise of the average density. Again, we produce 10,000 synthetic populations based on the distribution of Supplementary Fig.~\ref{fig:cpt_distrib}. Based on the dispersion of the synthetic YSO populations, we estimate more than 6$\sigma$ at the longitude resonance.

\subsection{External associations}
\label{sect:external_assoc}

The lack of a general and complete catalog of protostars exhibiting jets and the difficulty in obtaining all the parameters of interest from these objects are the most important limitations to definitely characterize the type and quantity of protostars in our Galaxy emitting $\gamma$ rays. The utilized RMS survey has been demonstrated as a useful tool for confirming the existence of GLPs, but only the brightest sources are included there. Up to date, in the literature, there are some well-known protostellar jets driven by massive YSOs that have been proposed as possible contributors to CR acceleration due to their proximity with a 4FGL source (e.g. \cite{2023Ruizhi, deOnaWilhelmi2023, MendezGallego2025, 2022Yan}). While HH 219 (IRAS 08211-4158) \cite{Araya2022} has already been studied as posible $\gamma$-ray emitters,
some of these objects are not included in our initial sample of GLPs because many of them are not compiled in the RMS catalog, and some others are highly affected by systematics in the 4FGL catalog and need a dedicated analysis to correct them. S255 NIRS3 may serve as an example of a potential $\gamma$-ray source that is not included in the RMS survey, whose hint of $\sim$3$\sigma$ variability was already studied in $\gamma$ rays correlating with a thermal outburst (see \cite{2017Moscadelli, 2023Fedriani, deOnaWilhelmi2023} for further details). Another prominent example is the well-studied  HH\,80--81 protostellar system (whose corresponding 4FGL source is shifted due to systematics; see \cite{MendezGallego2025}), or the nearby source HH\, 211 \cite{2023Ray}. In all cases, the association could not be firmly established. 
Additionally, recent observations using the James Webb Space Telescope have confirmed that even low-mass YSOs can drive collimated jets traveling at supersonic velocities through high-density mediums, producing shocks that may accelerate particles, such as the case of HH\,46--47 \cite{2024Nisini},    
where a $\gamma$-ray source (4FGL J0826.1-5053) is closely located. 

%\textcolor{red}{The case of  G030.1981-00.1691  is remarkable due to its proximity to a protoplanetary nebula. G030.1981-00.1691 is a massive YSO with clear signatures of driving a protostellar jet \cite{2024Ortega}. However, deep analyses of 4FGL J1846.9-0227 points toward another potential origin of the $\gamma$ rays in the protoplanetary nebula IRAS 18443-0231 \cite{2025Cala}. In this second case, leptonic emission may be present givent the fact that IRAS 18443-0231 is bright in X-rays. Despite both YSO and planetary nebula are contained inside the 68\% confident contour of the \emph{Fermi} source and the observed $\gamma$-ray luminosity seems achievable by the relation presented in Eq. (\ref{eq:Lgamma_vs_Lbol}), we can only conclude that this $\gamma$-ray emission may be affected by another possible origin rather than just protostellar jets.}

4FGL J1846.9-0227 has been previously studied an identified with IRAS 18443-0231 \cite{2024Ortega}, a protoplanetary nebula with non-thermal radio emission \cite{2025Cala}. The previous analysis focuses on the presence of non-thermal radio emission and X-ray emission arising from the protoplanetary nebula. However, in the same study, G030.1981-00.1691 is presented as a massive YSO driving a protostellar jet that is included in the 68\% confident area of the \emph{Fermi} source, remaining as an alternative counterpart for the detected $\gamma$ rays.

On the other hand, we observe the existence of some other protostellar radio jets separated by less than 0.5$^\circ$ from a $\gamma$-ray source that were not included in Table \ref{tab:GLPs}. Considering the limited capabilities of the \textit{Fermi}-LAT instrument for spatial resolution, we manually collected all external cases that we found with no other potential $\gamma$-ray emitters around. As a result, Table \ref{tab:literature_associations} shows the key parameters of these externally associated sources. Since we did not use these data points for performing the linear fitting presented in Eq. (\ref{eq:Lgamma_vs_Lbol}), external associations serve as proof of the potential extension of our finding to other non-associated objects by our methodology.

\section{Cloud density}
\label{sect:density}

The dominant radiative mechanisms considered for $\gamma$-ray production are relativistic Bremsstrahlung for electrons and pion decay for protons. Therefore, the ambient density in which accelerated particles interact with target material plays a crucial role in determining the $\gamma$-ray emission in star-forming environments. Although previous studies have examined the densities of molecular clouds (e.g. \cite{2009Butler, 2012Butler}), direct measurements of the ambient density in the vicinity of our targets are lacking. As a result, we estimate the density of the interstellar medium surrounding each GLP.

Adopting an average diffusion coefficient of $10^{28}$ $\rm cm^2 s^{-1}$ for 1 GeV of kinetic energy (see \cite{2007Strong} for further details), particles at this energy will travel tens of parsecs in $10^4$ yr. Therefore, we calculate the average density at large areas, which will be dominated by the molecular clouds where the protostars are born.

To calculate the ambient density around all our sources, we used the three-dimensional cubes from the $^{12}$CO($J=1\xrightarrow{} 0)$ maps  \cite{2001Dame}. 
We perform both spatial and spectral selection cuts of the cubes, considering the coordinates, the distance, and the radial velocity expected for each YSO. The spatial aperture where we integrate the velocity axis to obtain the CO intensity was a circle centered on the YSO position with the corresponding pixel radius to a projected length of 20~pc around the source. This radius is constrained to a maximum size of 5 pixels (8~pc in the most extreme case) for sources that are close enough and a minimum of 2 pixels for the furthest YSOs. Then, to compute the velocity cuts, we select an integration interval given by the standard deviation of a Gaussian fit to the spectral points related to the closest intensity peak regarding the radial velocity of our target, given in the RMS catalog. After the velocity integration, we correct our integration by a factor of $1/0.68$, and we applied a CO-to-H$_2$ conversion factor of $2\times 10^{20} {\rm cm^{-2}(K \, km \, s^{-1})^{-1}}$ \cite{2013Bolatto} to obtain the average H$_2$ column density in our spatial aperture. We multiplied our result by a factor of two to convert the H$_2$ density into proton density and assume a spherical volume with the same radius as the initial aperture to transform the column density into a three-dimensional density (see Supplementary Fig.~\ref{fig:density_calc}).

The resulting densities span from tens of particles per cubic centimeter to hundreds of them. The average density obtained from similar environments studied by \cite{2014Hou} is $\sim$150 $\rm cm^{-3}$, validating our results. Only for the case of G274.0649--01.1460A, we did not find any CO molecular cloud at the corresponding coordinates and radial velocity. Therefore, we used a density value of $\sim$70 $\rm cm^{-3}$ extracted from \cite{2022Kumar}. 

Due to all the intrinsic uncertainties in the cloud selection, we did not compute the errors of the density calculation in this work. Therefore, the y-axis error bars are underestimated, and the hidden uncertainties in the density could be the main contributors to the high dispersion shown Fig. \ref{fig:Lgamma_vs_Lbol}. We consequently took this into account while performing the linear fit of Fig. \ref{fig:Lgamma_vs_Lbol} (see \hyperref[sect:linear_fit]{Linear fitting} section).

The final comparison of the density calculated in the region as a function of the $\gamma$-ray and the bolometric luminosities is shown in Extended Fig.~\textcolor{blue}{2}. We can see that there is no correlation between none of the depicted luminosities and the density calculated in the region.

\subsection{Impact of density dispersion}

Since the hidden uncertainties of the cloud density estimation could affect to the correlation shown in Fig. \ref{fig:Lgamma_vs_Lbol}, we tested their potential impact in the correlation coefficient and thus the dispersion of data points. The uncertainties of the density can arise from the systematic uncertainties of the $^{12}$CO($J=1\xrightarrow{} 0)$ line emission, the velocity cuts, and the spatial cuts.

Firstly, to estimate the systematics associated with the CO emission, we produce a set of 1,000 mock CO maps with Gaussian dispersions given by the uncertainties of the CO composite survey (Table 1 of \cite{2001Dame}). Note that the systematic uncertainty depends on which of the 37 individual surveys cover the location of a given GLP. When a GLP is included in more than one individual survey, we adopt the lowest uncertainty among them. We find that, on average, the systematic errors are of the order of $\sim$1 cm$^{-3}$, a value we expect to be negligible compared with the other sources of uncertainty.

Regarding the other two possible sources of error, we find no satisfactory way to reliably estimate their magnitudes. Velocity cuts are highly dependent on the presence of other molecular clouds along the line of sight. The most critical situation arises when the targeted molecular cloud has a complex morphology, making it impossible to disentangle whether broad velocity distributions (even with multiple peaks) belong to a single structure or result from several independent clouds. In such cases, the error associated with the velocity cuts may be significant. The spatial–cut uncertainty is even more difficult to assess. We are roughly assuming that CRs propagate, on average, over scales of $\sim$20 pc, although the diffusion coefficient may be suppressed, which would reduce this scale substantially. In addition, we assume spherical symmetry when converting two-dimensional column densities into volumetric densities.

To evaluate whether uncertainties in density could bias the observed relationship between $L_\gamma$ and $L_{\rm bol}$, we test the impact of a random density distribution independent of any CO information. The density distribution of molecular clouds in the Milky Way can be inferred from the giant molecular cloud catalog compiled by \cite{2014Hou}. From this catalog, we obtain a broad (log-)normal distribution of the form $10^{2.5\pm1.0}$ cm$^{-3}$. We then randomly assign density values from this distribution to our GLPs and generate a sample of 100,000 realizations analogous to Fig. \ref{fig:Lgamma_vs_Lbol}, with the key parameters shown in Supplementary Fig.~\ref{fig:density_distrib}.

Supplementary Fig.~\ref{fig:density_distrib} show that, under random density distributions, we consistently recover a positive slope, with the results lying far from flat or negative correlations. This is clearly illustrated in the Pearson’s coefficient panel, where most of the mock datasets exhibit moderate positive correlations.  
Moreover, since none of the Pearson coefficients are negative and only $\sim$30\% fall below 0.5, we conclude that it is highly improbable that density uncertainties could hide the true densities in a way that artificially produces a correlation between $L_\gamma$ and $L_{\rm bol}$. Among our 100,000 mock density datasets, we did not find a single such case.

We also note that the observed dispersion of Fig. \ref{fig:Lgamma_vs_Lbol} is clearly lower than the average dispersion for random densities, indicating that the calculated densities for our 33 data points are following a physical behaviour. However, for computing the linear fitting is necessary to take into account the hidden dispersion that errors in $\rho$ can produce.

\section{Linear fitting}
\label{sect:linear_fit}

To fit the exponential relation found between $L_{\rm bol}$ and the cosmic ray abundance, we perform a linear Bayesian regression (e.g. \cite{2010Hogg}) to the logarithmic quantities shown in Fig. \ref{fig:Lgamma_vs_Lbol}. We, thus, launch a Markov chain Monte Carlo (MCMC) algorithm formed by 64 walkers with 10,000 steps each, including the 300 burnt-in steps and a thin value of 25 (approximately half of the autocorrelation time). As a result, we maximize the posterior probability, which depends on three parameters: the slope ($\alpha$), or the exponential index in Eq. (\ref{eq:Lgamma_vs_Lbol}); the intercept ($\beta$) indicating the constant to transform $(L_{\rm bol})^\alpha $ into $(L_\gamma \, \rho^{-1} )$; and the fractional intrinsic dispersion factor ($f$), which will also absorb the dispersion caused by the error in the density calculations (see \hyperref[sect:density]{Cloud density} section). To do so, we evaluate the following likelihood distribution:

$$ \mathcal{L}_{\rm fit} \left( y_i \, | \, x_i, \sigma_{x, i}, \sigma_{y, i}, \alpha , \beta , f  \right) = \frac{1}{\sqrt{2 \pi s_i^2}} \exp \left[ - \frac{\left( y_i - \alpha x_i - \beta \right) ^2}{2s_i^2}   \right] ,$$
with

$$ s_i^2 = \sigma_{x, i}^2 + \sigma_{y, i}^2 + f^2 \left( \alpha x_i + \beta \right) .$$

The equation above returns the probability of having the corresponding y-value for a given x-value, given uncertainties ($\sigma_{x, i}$ and $\sigma_{y, i}$), and a given set of parameters regulating our linear model and dispersion. We then multiply all the probabilities for the 33 GLPs in our sample, building the overall likelihood of the linear model. In addition, we use uninformative priors for the three parameters, limiting the slope in a uniform range of $(0, 3)$, the intercept in $(0, 100)$, and the fractional dispersion factor in $(10^{-10}, 10^1)$. Employing \texttt{emcee} \cite{2013Foreman}, we successfully fit the CR production versus the bolometric luminosity of our GLPs, as Supplementary Fig.~\ref{fig:corner} shows.

\vspace{5mm}

\section*{Extended items}

\begin{figure}[h]
    \centering
    \includegraphics[width=0.5\linewidth]{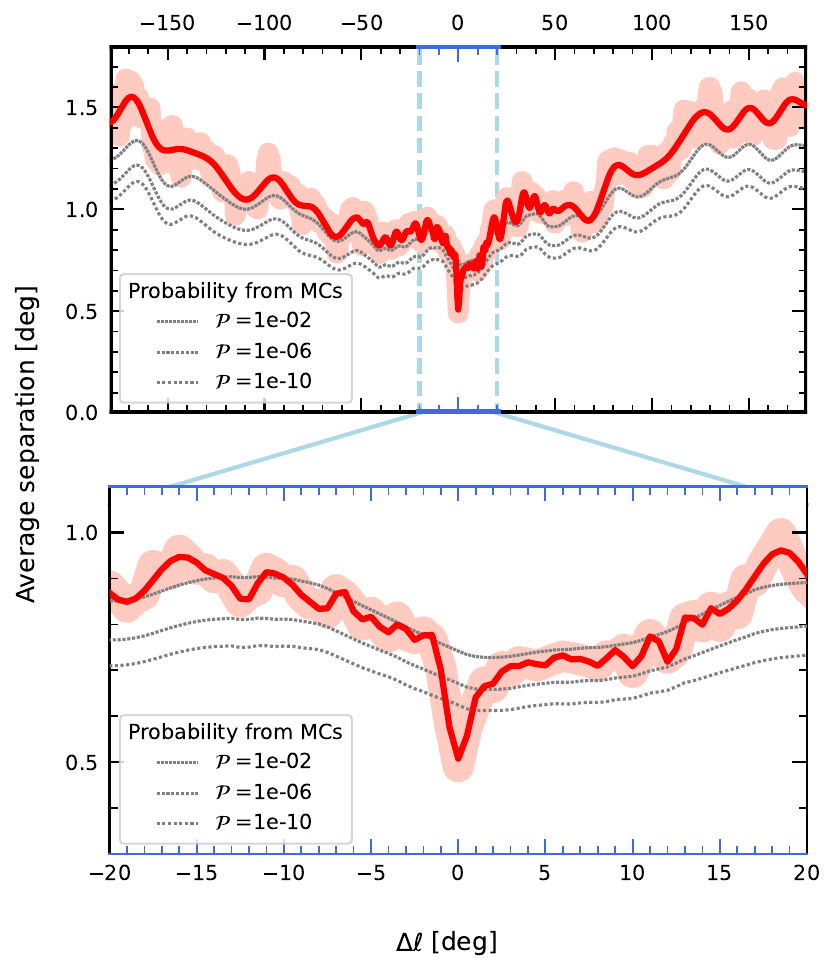}
    \caption{{Longitude resonance plot for the 4FGL and the closest YSOs counterparts}. The light red wide line shows the average separation for the real YSO population, whereas the red solid line is the smoothing result. Grey dashed lines represent the probability of obtaining the average separation values obtained from 10,000 synthetic populations.}
    \label{fig:hillas_shift}
\end{figure}
\clearpage

\begin{figure}[h!]
    \centering
    \includegraphics[width=0.7\linewidth]{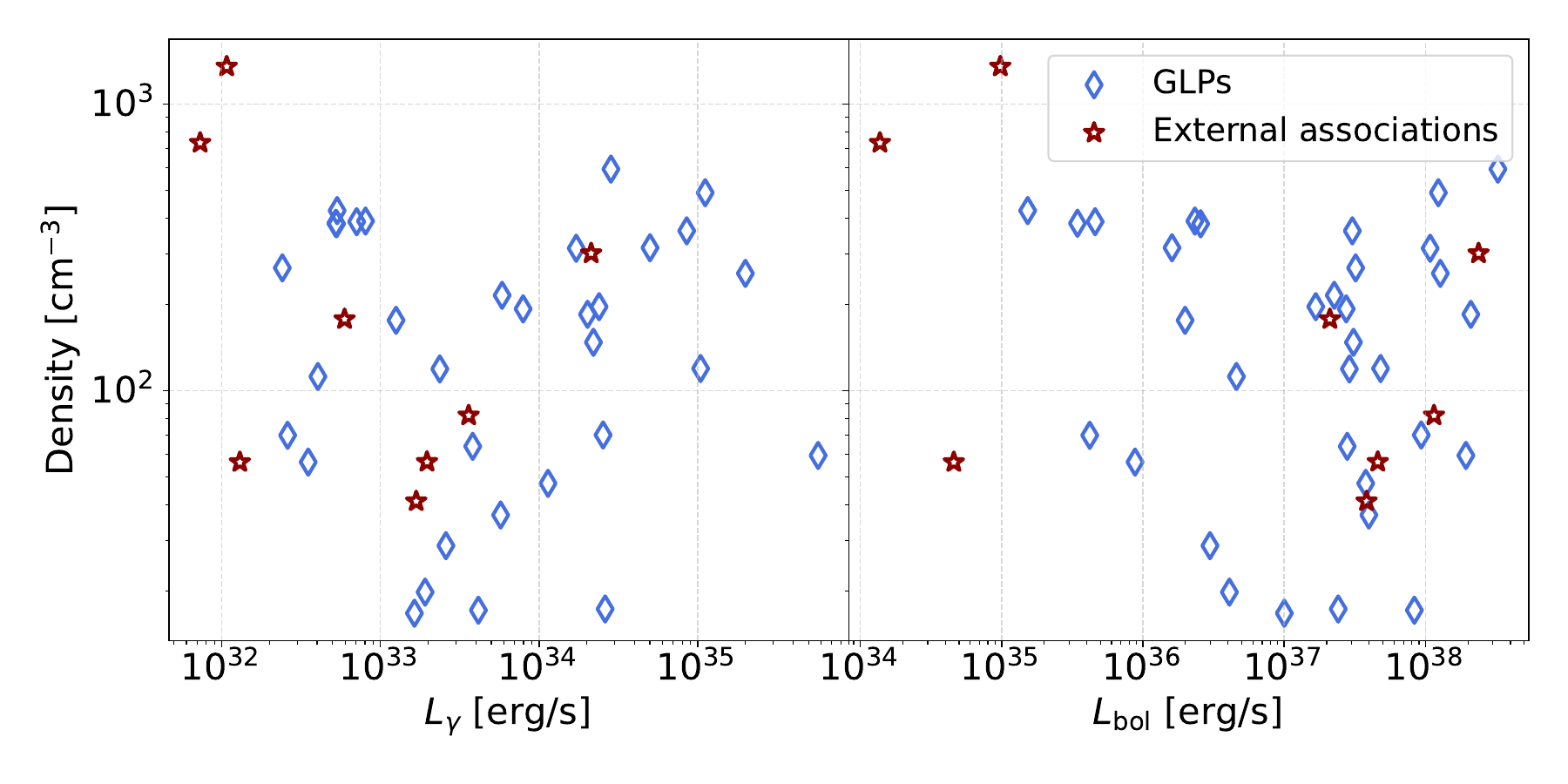}
    \caption{Ambient density as a function of source luminosities. Left: Density computed for likelihood-ratio associations and external sources (Table \ref{tab:GLPs} \& \ref{tab:literature_associations}, respectively) versus $\gamma$-ray luminosity. Right: Same densities versus bolometric luminosities of YSOs. No correlation is observed.}
    \label{fig:Lgamma_vs_density}
\end{figure}

\begin{figure}[h]
    \centering
    \captionof{table}{Positional coincidences depending on the size of the elliptical area for accepting associations. Significance of the association along with false association form synthetic populations is reported.}
    \includegraphics[width=0.77\linewidth]{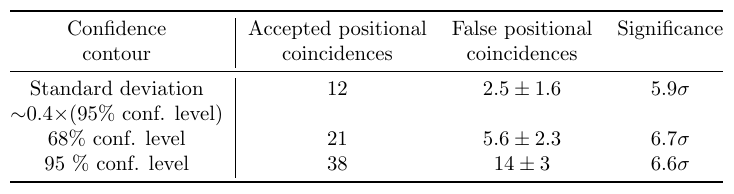}
    \label{tab:positional_coincidences}
\end{figure}

%%%%%%%%%%%%%%%% ACKNOWLEDGEMENTS %%%%%%%%%%%%%%%

\vspace{5mm}

\noindent {\bf Data availability: } 
All data used in this work are publicly available. The RMS survey is accessible via VizieR at \url{http://vizier.cds.unistra.fr/viz-bin/VizieR?-source=J/ApJS/208/11}. The \textit{Fermi}-LAT 4FGL-DR4 catalog can be downloaded from \url{https://fermi.gsfc.nasa.gov/ssc/data/access/lat/14yr_catalog/}. A digitalized version of our results from Table \ref{tab:GLPs} are publicly available at: \url{https://github.com/VHEGA/glps.git}. Further information and extended analyses are available in the Supplementary Information.

\vspace{2mm}

\noindent {\bf Code availability: } 
Example of the code employed for statistical associations is publicly available at: \url{https://github.com/VHEGA/glps.git}.

\vspace{2mm}

\noindent {\bf Acknowledgments: }
This work makes use of data from the \textit{Fermi} Gamma-ray Space Telescope, a NASA mission, obtained from the Fermi Science Support Center.
We also made use of data products from the Midcourse Space Experiment. Processing of the data was funded by the Ballistic Missile Defense Organization with additional support from NASA Office of Space Science. This research has also made use of the NASA/ IPAC Infrared Science Archive, which is operated by the Jet Propulsion Laboratory, California Institute of Technology, under contract with the National Aeronautics and Space Administration.
This publication makes use of data products from the Two Micron All Sky Survey, which is a joint project of the University of Massachusetts and the Infrared Processing and Analysis Center/California Institute of Technology, funded by the National Aeronautics and Space Administration and the National Science Foundation.
This work also uses data products from the Herschel Instrument. Herschel is an ESA space observatory with science instruments provided by European-led Principal Investigator consortia and with important participation from NASA.
J.M.-G. acknowledges lessons and materials giving by Dr. Jean Ballet during the First CTAO School held in Bertinoro (Italy) in July 2024.
J.M.-G. also acknowledges financial support from the FPI-Severo Ochoa grant CEX2021-001131-S-20-6, PRE2022-103386 funded by MICIU/AEI/ 10.13039/501100011033 and ESF+.
J.M.-G., R.L.-C., S.M. and I.A. also acknowledge financial support from the Spanish "Ministerio de Ciencia e Innovaci\'on" through grant PID2022-139117NB-C44. 
J.M.-G., R.L.-C., S.M., R.F. and I.A. acknowledge financial support from the Severo Ochoa grant CEX2021-001131-S funded by MCIN/AEI/ 10.13039/501100011033.
R. L.-C. and S.M. also acknowledge the financial support through grant CNS2023-144504 funded by MICIU/AEI/ 10.13039/501100011033 and by the European Union NextGenerationEU/PRTR. 
EdOW acknowledges the support of DESY (Zeuthen), a member of the Helmholtz Association HGF.
R.F. acknowledges financial support from PID2023-146295NB-I00.

\vspace{2mm}

\noindent {\bf Author Contribution Statement: } 
J.M.-G., R.L.-C., E.O.W., S.M., I.A., and R.F. have contributed to the scientific discussions and interpretations covered in this work. J.M.-G., R.L.-C., and E.O.W. have contributed to the writing of this manuscript. J.M.-G. has lead the majority of the scientific methodology, including the association mechanisms, the correlation and fitting procedures, and the investigation of counterparts, supervised by R.L.-C., E.O.W, and S.M. J.M.-G., R.L.-C., and S.M. have contributed to the development of the main plots. R.L.-C. and E.O.W. have conducted the discussion regarding the hadronic nature of $\gamma$-ray emission and acceleration site, helped by J.M.-G
\vspace{2mm}

\noindent {\bf Competing interests: }
The authors declare no competing interests.

\setcounter{figure}{0}
\setcounter{table}{0}

\renewcommand{\figurename}{Extended Fig.}
\renewcommand{\tablename}{Extended Table}

%%==================================%%
%%        Supplementary info        %%
%%==================================%%

\section*{Supplementary Information}
\ \\ 
Supplementary Sections A -- F \\
Supplementary Figures 1 -- 20 \\
Supplementary Tables 1 \& 2 \\
Supplementary References [1 -- 57] \\

%%==================================%%
%%            References            %%
%%==================================%%

%\bibliography{references}

%%==================================%%
%%          Supplementary           %%
%%==================================%%
\clearpage
\appendix

\renewcommand{\thepage}{S\arabic{page}}
\setcounter{page}{1}

\renewcommand{\thefigure}{\arabic{figure}}
\renewcommand{\figurename}{Supplementary Fig.}
\renewcommand{\tablename}{Supplementary Table}
\renewcommand{\thetable}{\arabic{table}}
\setcounter{figure}{0}
\setcounter{table}{0}
%\setcounter{equation}{0}

%%===============================%%
%%      Supplementary title      %%
%%===============================%%
\vspace*{10mm}
\begin{center}
\huge
    {\bf Supplementary Information of:}
    \vspace{4mm}
    
    Evidence for protostellar jets as a population of hadronic gamma-ray sources
\end{center}

\vspace{2mm}
\begin{center}
\Large
    Javier~M\'endez-Gallego$^{1\, \ast}$, Rub\'en~L\'opez-Coto$^{1}$, Emma~de~O\~na~Wilhelmi$^{2}$, Stefano~Menchiari$^{1}$, Iv\'an~Agudo$^{1}$, Rub\'en~Fedriani$^{1}$
\end{center}

\vspace{1mm}
\begin{center}
\large
    $^1$Instituto de Astrof\'isica de Andaluc\'ia (IAA), CSIC, Glorieta de la Astronom\'ia s/n, Granada, 18008, Spain.
    
    $^2$Deutsches Elektronen-Synchrotron, DESY, Platanenallee, 6, Zeuthen, 15738, Germany.

    \vspace{3mm}
    $^\ast$Corresponding author(s). E-mail(s): \url{jmendez@iaa.es}

    Contributing authors: \url{rlopezcoto@iaa.es};
    
    \url{emma.de.ona.wilhelmi@desy.de}
\end{center}

\clearpage

\section{Supplementary tables and figures}

In these section, we present tables and figures used along the \hyperref[sect:methods]{Methods} sections.

\begin{figure}[h]
    \centering
    \captionof{table}{Number of sources included in the catalogs used for $\gamma$-rays and YSOs. Total amount of sources are compared to the final filtered sources of interest.}
    \includegraphics[width=0.49\linewidth]{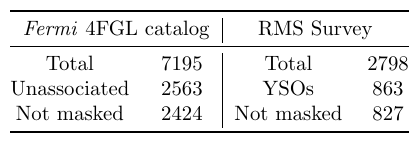}
    \label{tab:catalogs}
\end{figure}

\begin{figure}[h]
    \centering
    \includegraphics[width=0.8\linewidth]{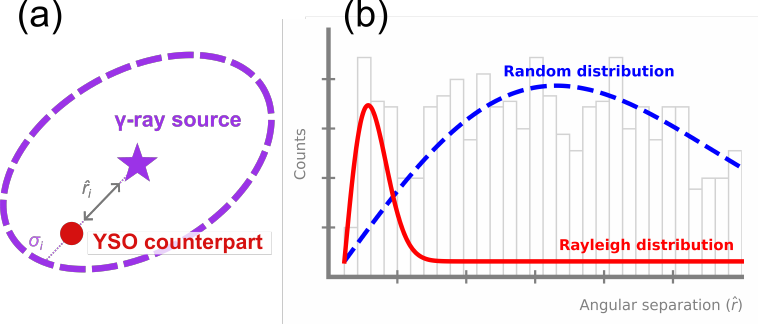}
    \caption{Illustration of the initial state for the association mechanism. \textit{Panel (a)} shows a $\gamma$-ray source with its positional uncertainty shown in purple, along with the closest YSO represented as a red dot, indicating it is treated as a point-like source. This serves as the starting point, where we begin from given values for $\hat{r}$, $\sigma$, and $\rho_{\rm cpt}$ to distinguish true associations from random ones. \textit{Panel (b)} presents both likelihood distributions, highlighting their differences. The background histogram presents a random sample of sources composed of both Rayleigh and random distributions.}
    \label{fig:assoc_scheme}
\end{figure}

\begin{figure}[h]
    \centering
    \includegraphics[width=0.8\linewidth]{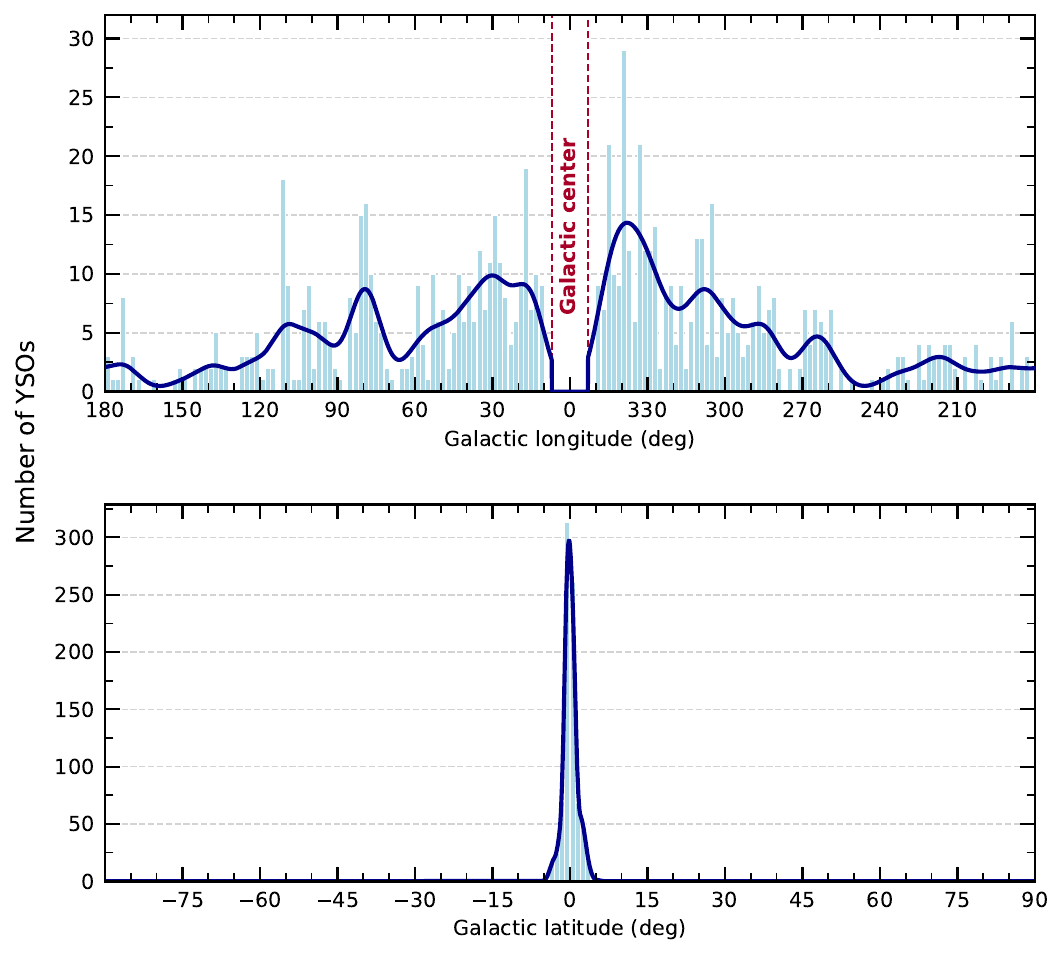}
    \caption{{Marginal distributions of YSOs from the RMS Survey along the Galactic coordinates}. Red dashed lines mark the masked region where the survey is biased because of the proximity to the Galactic center.}
    \label{fig:cpt_distrib}
\end{figure}

\begin{figure}[h]
    \centering
    \includegraphics[width=0.8\linewidth]{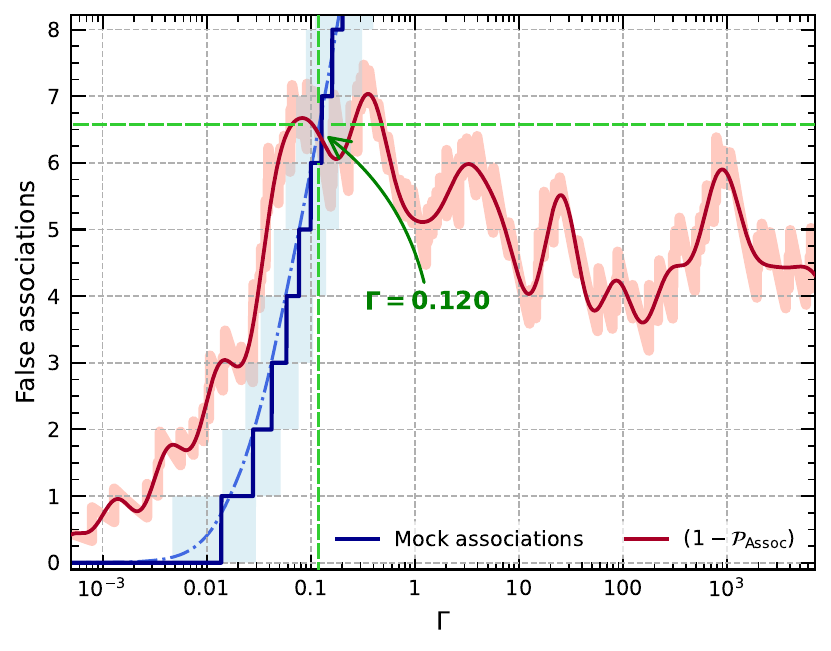}
    \caption{{False associations among the accepted associations versus the prior ratio}. 
    Red colors represent the false associations obtained from summing all the $(1- \mathcal{P}_{\rm Assoc})$ for the selected associations where $\mathcal{P}_{\rm Assoc} >0.5$. The solid red line represents a softening of the light-red wide line in the background. 
    Blue solid line indicates the average number of accepted associations with $\mathcal{P}_{\rm Assoc} >0.5$ obtained from the synthetic sample of 10,000 mock populations. The shaded area is the 1$\sigma$ error of this MC estimation for the false associations. The dash-dotted line is the summation of the Poisson probabilities of finding a YSO closer than the separation observed between each pair of matched sources, which is compatible with the accepted associations from the mock populations.}
    \label{fig:Gamma}
\end{figure}

\begin{figure}[h!]
    \centering
    \includegraphics[width=0.8\linewidth]{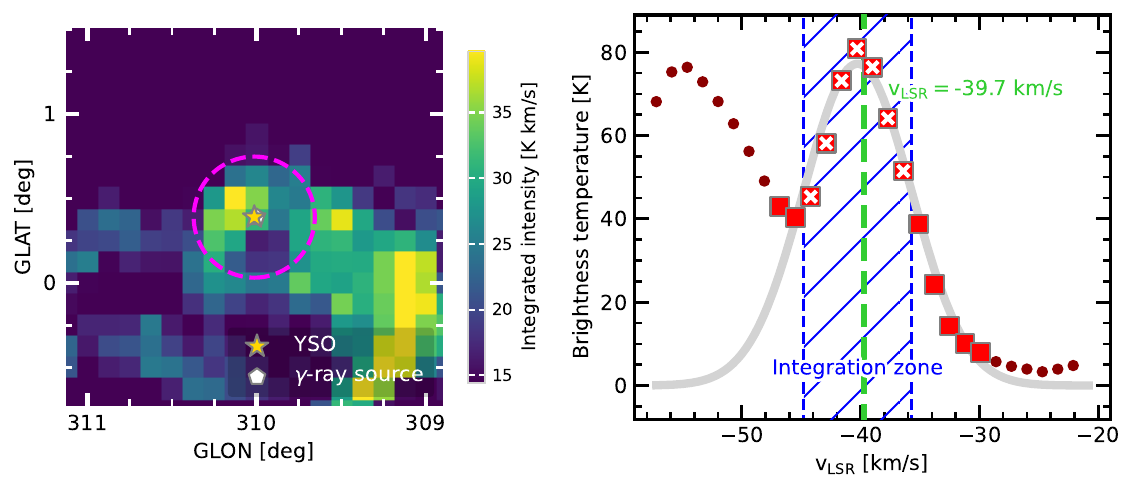}
    \caption{$^{12}$CO($J=1\xrightarrow{} 0)$ data cube used to estimate the ambient density around G310.0135+00.3892 (IRAS 13481–6124), yielding $\rho \simeq 180\ \mathrm{cm^{-3}}$. Left: moment-0 map with the analyzed region (magenta circle) at the selected velocity cuts. Right: velocity profile with a Gaussian fit (grey), performed using only the data points highlighted by red squares, selected from all flux measurements (dark red circles) around the peak closest to the desired $v_{\rm LSR}$. The source velocity is indicated by a green dashed line, and the integration range is shown in blue, with white crosses marking the points used for the integrated flux. A separate CO component at $-56\ \mathrm{km\ s^{-1}}$ is also visible. 
    }
    \label{fig:density_calc}
\end{figure}

\begin{figure}[h!]
    \centering
    \includegraphics[width=0.95\linewidth]{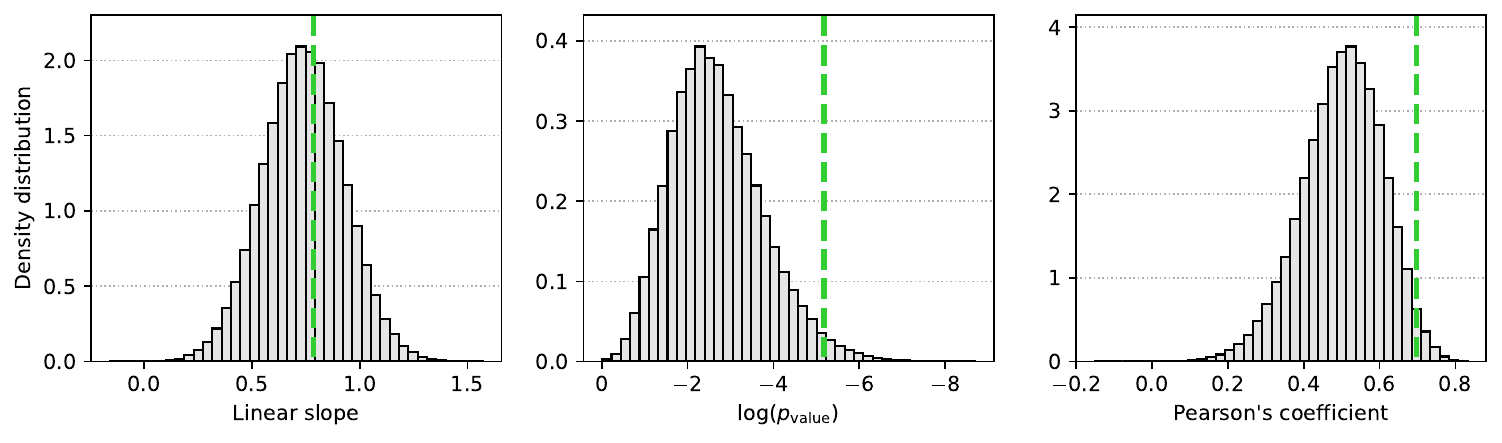}
    \caption{{Dispersion of the $L_{\rm bol}$ -- $L_\gamma /\rho$ correlation parameters with random density dispersion.} Green vertical lines represent the parameters found in this work: \textit{Slope} $=0.79$ (left); $p_{\rm val}=6.6\times 10^{-6}$ (middle); and \textit{Pearson's} $=0.70$ (right).}
    \label{fig:density_distrib}
\end{figure}

\begin{figure}[h]
    \centering
    \includegraphics[width=0.8\linewidth]{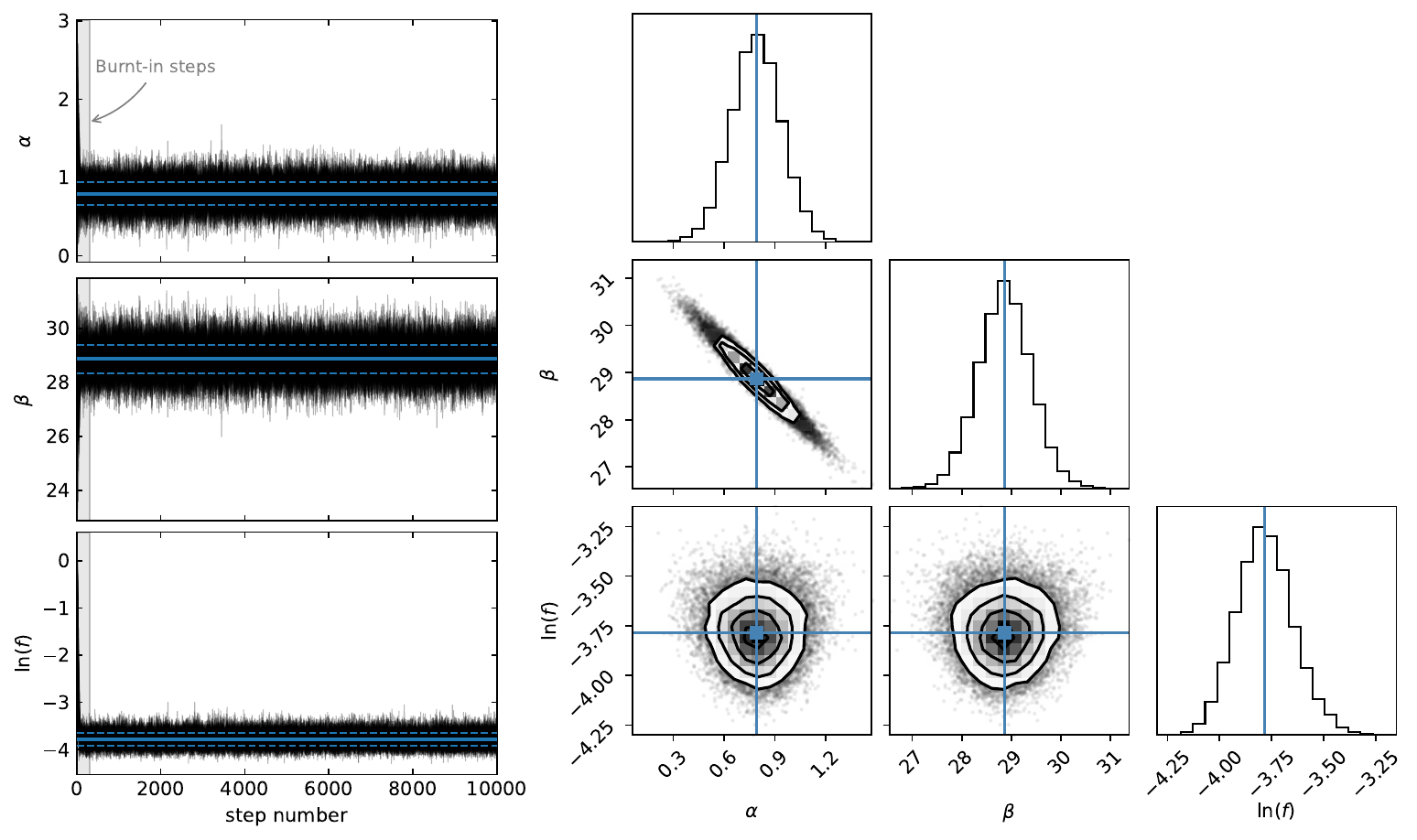}
    \caption{{MCMC resulting chains for optimizing the linear regression of $L_{\rm bol}$ -- $L_\gamma /\rho$ correlation}. While black colors represent the values for the three parameters of the regression in each step, blue colors highlight the resulting values. Grey shaded regions mark the burnt-in steps. An important dependency of the intercept is observed when compared to the slope: the steeper relation ($\alpha$), the lower intercept ($\beta$). The intrinsic dispersion ($\ln f$) is statistically uncorrelated with the rest of the model parameters.}
    \label{fig:corner}
\end{figure}

\clearpage

\section{Nature of the particle population and acceleration site}
\label{sect:kinetics}

\begin{figure}[h!]
    \centering
    \includegraphics[width=0.72\linewidth]{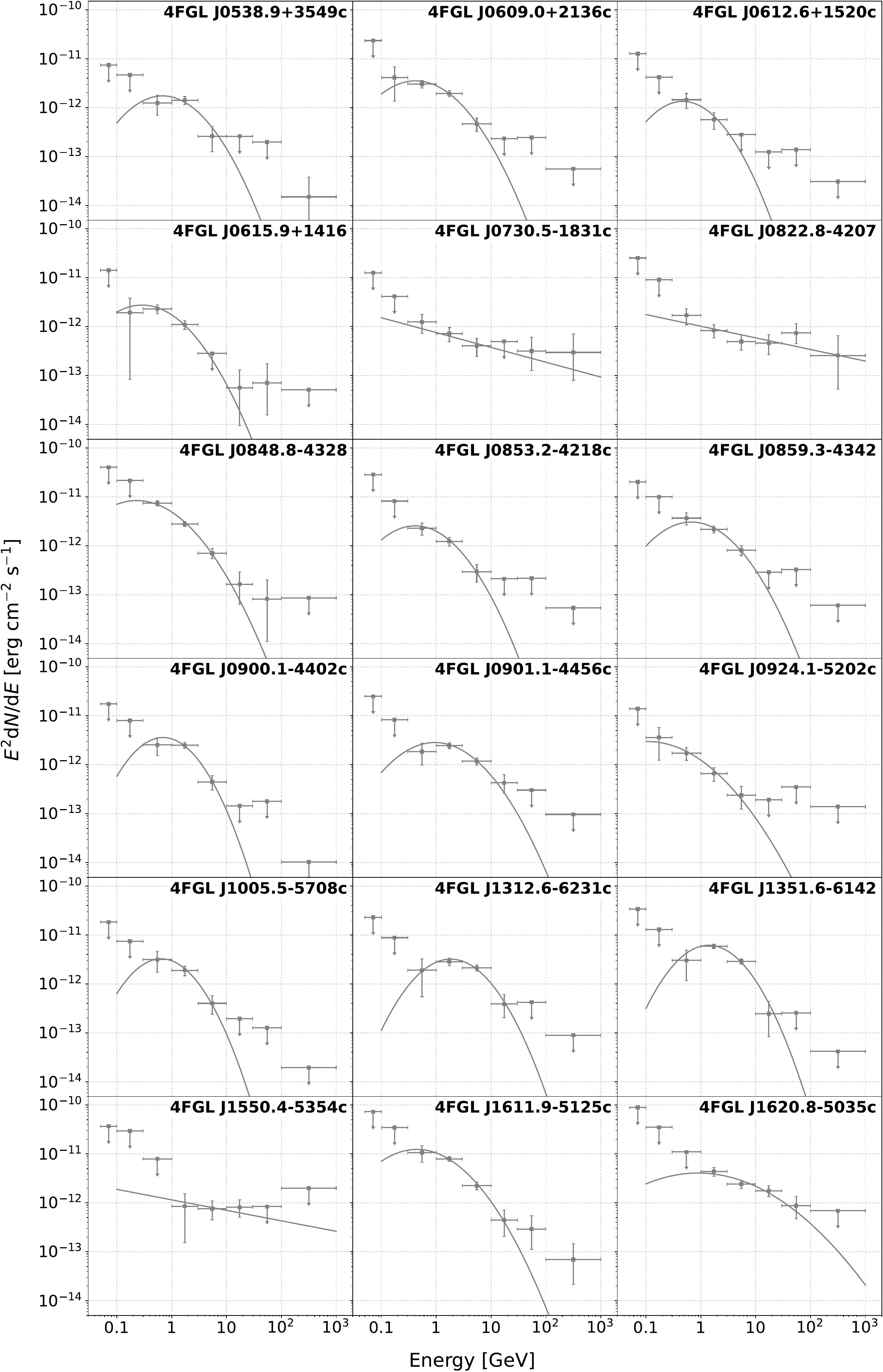}
    \caption{4FGL \emph{Fermi}-LAT spectral points for 18 out of the 33 identified GLPs. Flux points and their corresponding errorbars are extracted from the 4FGL-DR4 catalog, as well as the best fit spectra (black solid lines). X-axis errorbars show the energy bins, while Y-axis illustrate the flux uncertainties in each range.}
    \label{fig:fermi_spectra}
\end{figure}

\begin{figure}[h!]
    %\ContinuedFloat
    \centering
    \includegraphics[width=0.72\linewidth]{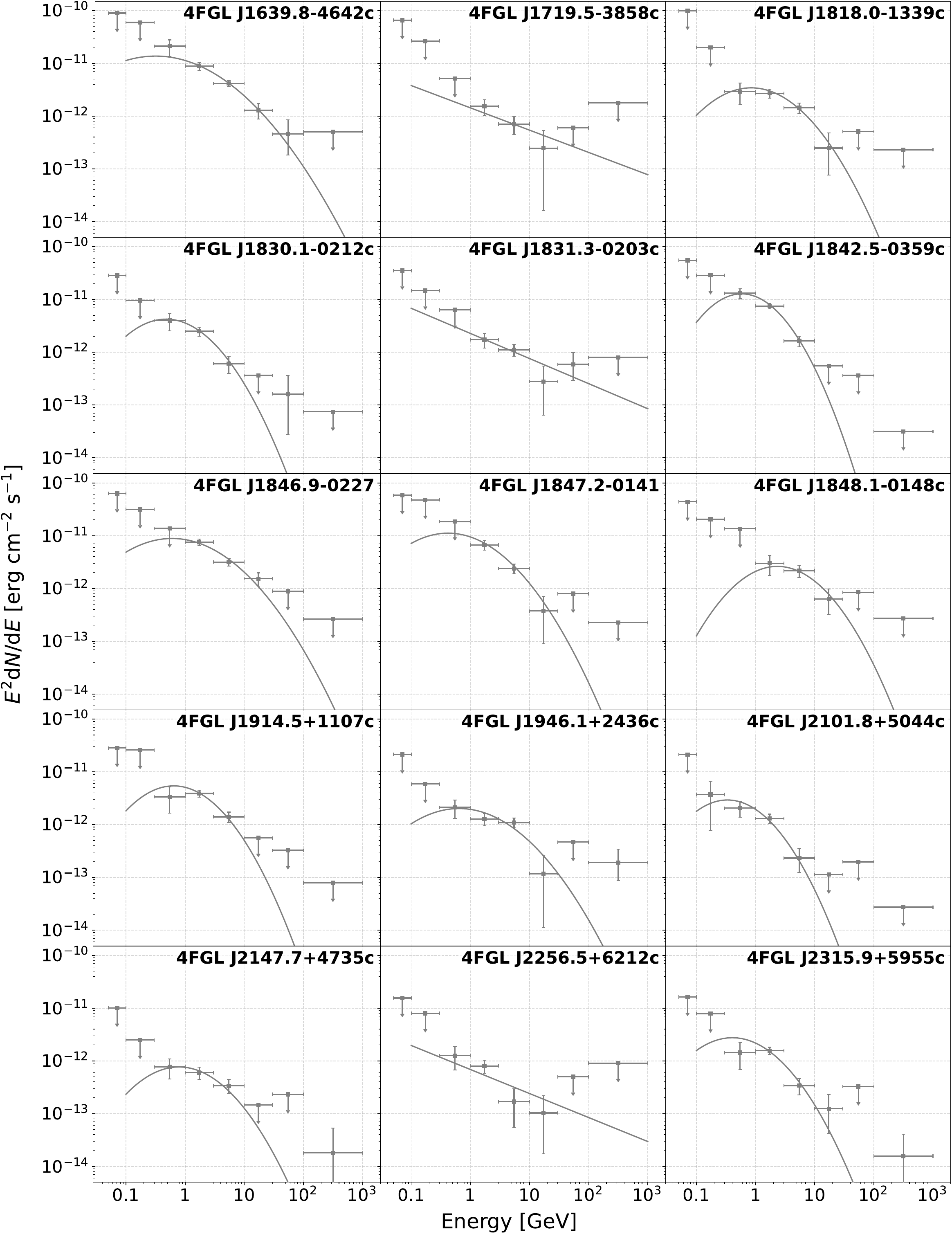}
    \caption{4FGL \emph{Fermi}-LAT spectral points for the remaining 15 of the 33 identified GLPs. Flux points and their corresponding errorbars are extracted from the 4FGL-DR4 catalog, as well as the best fit spectra (black solid lines). X-axis errorbars show the energy bins, while Y-axis illustrate the flux uncertainties in each range.}
    \label{fig:fermi_spectra2}
\end{figure}

To discern the nature of the particle population responsible of the $\gamma$ ray emission, we evaluate the total kinetic and maximum energy observed of such particles, and put it in context of the two radiation mechanisms that can account for the high energy photons: $\pi^0$ decay for protons and Bremsstrahlung for relativistic electrons (see Fig. 4 of \cite{deOnaWilhelmi2023s} and Fig. 6 of \cite{MendezGallego2025s}). The third channel typically involved in $\gamma$-ray emission, inverse Compton scattering, is generally negligible in the environment of protostars due to the relatively low ambient photon density compared to the high particle density where star-formations occurs.

Supplementary Figs. \ref{fig:fermi_spectra} \& \ref{fig:fermi_spectra2} illustrate the $\gamma$-ray spectrum provided by the \textit{Fermi} 4FGL catalog for the selected sources compiled in Table \ref{tab:GLPs}. 
To obtain the distribution of the particle population, we fit the $\gamma$-ray spectrum using the software \texttt{Gamera} \cite{Gamera_2022s, 2015ICRC...34..917Hs} in which we can introduce a particle spectrum with an arbitrary spectral shape, make it interact with a medium of density $\rho$ and calculate what is the resulting photon spectrum. 
Our interests lie on calculating the maximum energies that protons or electrons must reach to reproduce the observed $\gamma$-ray spectra and the integrated non-thermal kinetic energy of the particle distribution. We note that our associated $\gamma$-ray sources show soft spectral indexes ($\alpha>2$), highlighting the importance of the low energy break to integrate the full kinetic energy contained in the particle distribution.  Therefore, we fitted the $\gamma$-ray spectrum by injecting a particle distribution in momentum space~($p$) --instead of energy-- of the form of a power law with an exponential cut-off ($p_{\rm cut}$) [see Eq. (\ref{eq:fp})]. When transforming momentum to energy, our particle distribution will then naturally set a low energy break.

\begin{equation}
    f(p) = \frac{f_0}{\rm (kg\, m\, s^{-1})^{3}} \left( \frac{p}{\rm kg \, m \, s^{-1}} \right)^{-(\alpha+2)} \exp \left( - \frac{p}{p_{\rm cut}}  \right),
    \label{eq:fp}
\end{equation}
where $p_{\rm cut}$ is the corresponding momentum of a particle with energy $E_{\rm cut}$, that is a free parameter as well as $f_0$ and  $\alpha$. In addition, we input the density values computed in \hyperref[sect:density]{Methods} and compiled in Table \ref{tab:GLPs}. Low $E_{\rm cut}$ ($\sim$1~GeV) implies difficulties to determine the power-law index ($\alpha$), limited to a minimum of 1.

\begin{figure}[h!]
    \centering
    \includegraphics[width=0.8\textwidth]{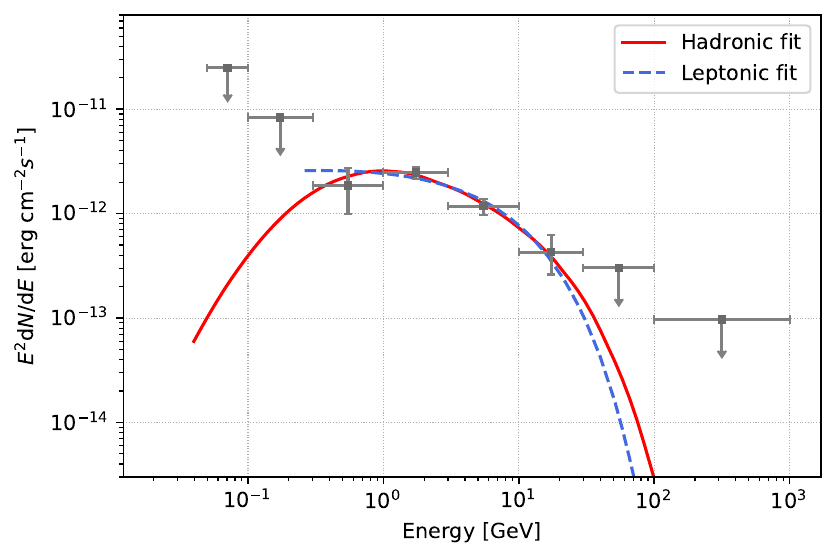}
    \caption{Example of a  \texttt{Gamera} fit to the \emph{Fermi}-LAT spectral points of 4FGL J0901.1-4456c. Spectral points and uncertainties extracted from the 4FGL-DR4 catalog are shown in grey. Hadronic fit (red solid line): $f_0=10^{69\pm 14}$, $\alpha=1.8\pm 0.7$, $E_{\rm cut}=10^{1.62 \pm 0.23}$~GeV. Leptonic fit (blue dashed line): $f_0=10^{64\pm 22}$, $\alpha=2.0\pm 1.2$, $E_{\rm cut}=10^{1.14\pm 0.22}$ GeV.}
    \label{fig:hadronic_leptonic_fit}
\end{figure}

 The cut-off energy ($E_{\rm cut}$) is defined by the maximum energy achieved by the accelerated particles. 
 We would like to note that this is the minimum cut-off energy compatible with the performed fit, but higher cut-off energies would be possible if higher energy spectra were measured. The lower end of the {\it Fermi}-LAT photon spectrum is usually represented by upper limits. In general, to avoid overshooting them, the power-law expression above has to be modified with a low-energy cutoff to achieve convergence on the fit. However, these low energies should be carefully treated due to the limited angular and energy resolution of the Fermi-LAT below 200~MeV energies. For the purpose of this paper, we choose not to include the low-energy upper limits in the fitting procedure, ignoring the potential low-energy cutoff and instead focusing on the higher-energy part of the spectrum, where the instrument response is better understood. An example of a leptonic and a hadronic fit is shown in Supplementary Fig. \ref{fig:hadronic_leptonic_fit}.

\begin{figure}[h!]
    \centering
    \includegraphics[width=0.7\linewidth]{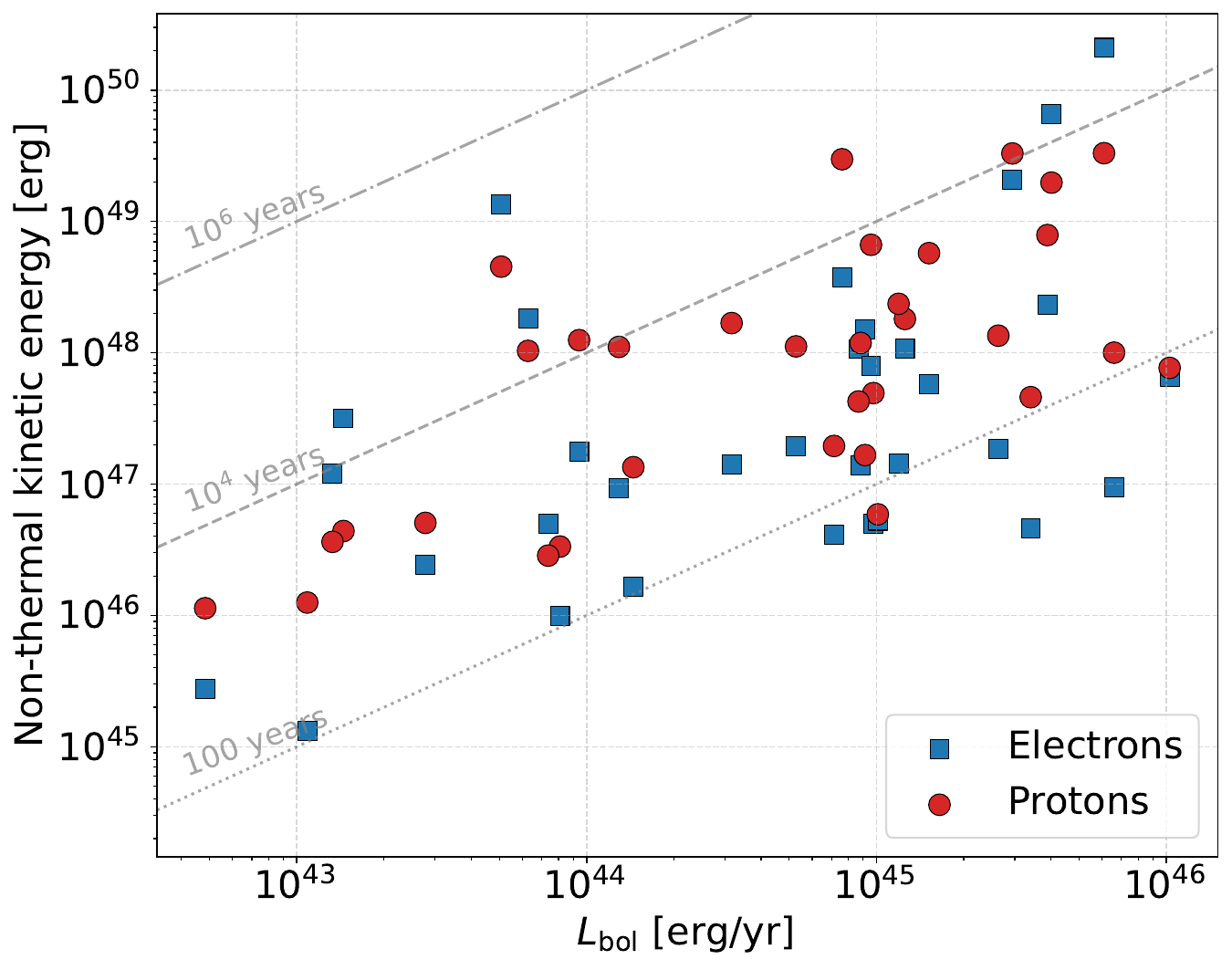}
    \caption{{Non-thermal kinetic energy obtained assuming a power-law spectrum for the particle distribution}. Grey dashed and dotted lines mark the accumulation timescales assuming that the 100\% of the available power as bolometric luminosity turns into non-thermal energy. Blue squares and red filled circles indicate electrons and protons, respectively}
    \label{fig:total_kinetik_energy_vs_Lbol}
\end{figure}

After performing the fits described above, we calculated the non-thermal kinetic energy by integrating the fitted particle spectra from a conservative 1 MeV. Since electron mass is significantly lighter than that for protons, the low energy break will be situated at $\sim$0.5 MeV, so the contribution of this low energy part of the particle distribution will be very significant in the leptonic case. The integrated results are compared to the bolometric luminosity of the protostellar objects in Supplementary Fig. \ref{fig:total_kinetik_energy_vs_Lbol}, used as a proxy of the jet mechanical power. We find that, if the entire energy budget of the system (here assumed to be $L_{\rm bol}$) were dedicated to particle acceleration, the required energy to accelerate protons and electrons could be accumulated within approximately 10$^4$ years in most of the cases.

However, in a more realistic scenario where the jet power is a portion of the observed bolometric luminosity, only a fraction of the mechanical power would be available for particle acceleration --typically assumed to be 1\% for electrons for high Mach-number shocks and 10\% for protons \cite{Bell2013s}. That would imply that, for under these conditions, the timescales required to accelerate protons remain compatible with the typical jet lifetimes, which are on the order of ten thousands to millions of years. In contrast, the timescales required for electrons to provide the energy budget necessary to explain the observed emission overcome the synchrotron cooling times of the hundreds-of-GeV electrons in the $\sim$0.1 mG magnetic fields. This argue against electrons as origin of the radiation observed and supports the hypothesis that protons are the dominant contributors to the $\gamma$-ray emission observed in GLPs.

\begin{figure}[h]
    \centering
    \captionof{table}{Maximum particle energy. Two acceleration sites are considered: surface accretion shocks and protostellar jet shocks.}
    \includegraphics[width=0.55\linewidth]{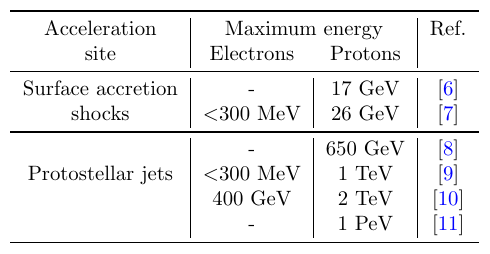}
    \label{tab:maximum_energy}
\end{figure}

Additionally, the maximum energy attained further supports the hadronic origin, excluding a pure leptonic Bremsstrahlung process. The role of relativistic electrons in protostellar jets has been extensively discussed in the literature, with their maximum energies always constrained to be lower than those of protons. Supplementary Table \ref{tab:maximum_energy} summarizes the maximum kinetic energies predicted before the discovery of GLPs, based on comprehensive models of electron and proton acceleration in both protostellar jet shocks and stellar surface accretion shocks. 

\begin{figure}[h]
    \centering
    \includegraphics[width=0.75\linewidth]{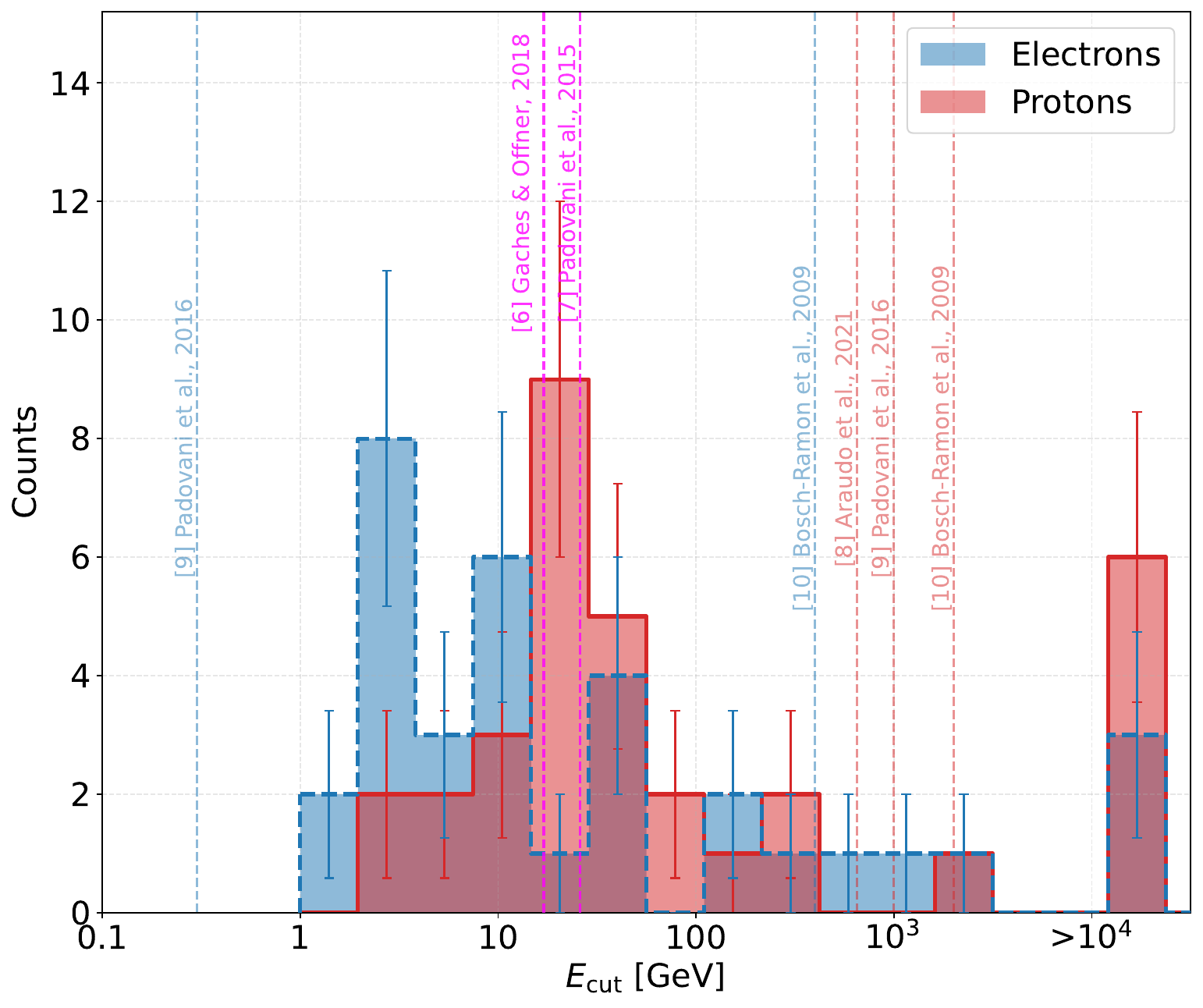}
    \caption{{Histograms of the maximum energies of electrons and protons. Poisson uncertainties are represented by error bars}. Vertical lines correspond to the maximum theoretical energies in the literature, magenta for those in surface accretion shocks, and in blue (red) those corresponding to electrons (protons) accelerated in the jet.}
    \label{fig:histogram_maximum_energies}
\end{figure}

The values of $E_{\rm cut}$ obtained for the electron/proton parent population fit to the data are shown in Supplementary Fig. \ref{fig:histogram_maximum_energies}. The maximum energies summarized in Supplementary Table \ref{tab:maximum_energy} are also added for comparison. In the case of electrons, the minimum energies to explain the 4FGL sources exceed the theoretical expectations for many of the GLPs, supporting hadronic scenarios.

We also investigate these maximum energies reachable by a protostellar jet based on simple (but solid) arguments, without considering complex theoretical models.  
The absolute maximum energy (for an ideal flow case) can be obtained by comparing the acceleration time ($t_{\rm acc}$) that particles need to reach such energy with their escape time ($t_{\rm esc}$) \cite{1984ARA&A..22..425Hs}:

$$t_{\rm acc} = \eta \frac{R_{\rm L}}{c};$$
$$t_{\rm esc} =\frac{R}{\beta c},$$
where $\eta=1/\beta^2=c^2/v^2$, $v$ the shock speed, $c$ the speed of light, $R_{\rm L}$ is the Larmor radius of the  particle and $R$ the radius of the acceleration site. Since the Larmor radius in the relativistic regime is given by:

$$R_{\rm L}=\frac{E}{q B c},$$
with $E$ being the energy of the particle, $q$ its charge and $B$ the magnetic field in the region. If we equal $t_{\rm acc}$ with $t_{\rm esc}$, we can derive the maximum energy $E_{\rm max}$ a particle can reach on a given region with magnetic field $B$:

$$E_{\rm max}= q B R v=3\left(\frac{v}{10^3 \rm{km/s}}\right) \left(\frac{R}{ \rm{pc}}\right) \left(\frac{B}{ \mu\rm{G}}\right) \rm{TeV},$$
which is the well-known Hillas equation. The maximum energy is thus proportional to the shock speed, the radius of the acceleration site, and the magnetic field. A representation of the maximum energy that can be achieved in a shock with radius $R=10^{16}$ cm as a function of the shock speed and magnetic field is shown in Supplementary Fig. \ref{fig:hillas_maximum_energy}.

\begin{figure}[h]
    \centering
    \includegraphics[width=0.7\linewidth]{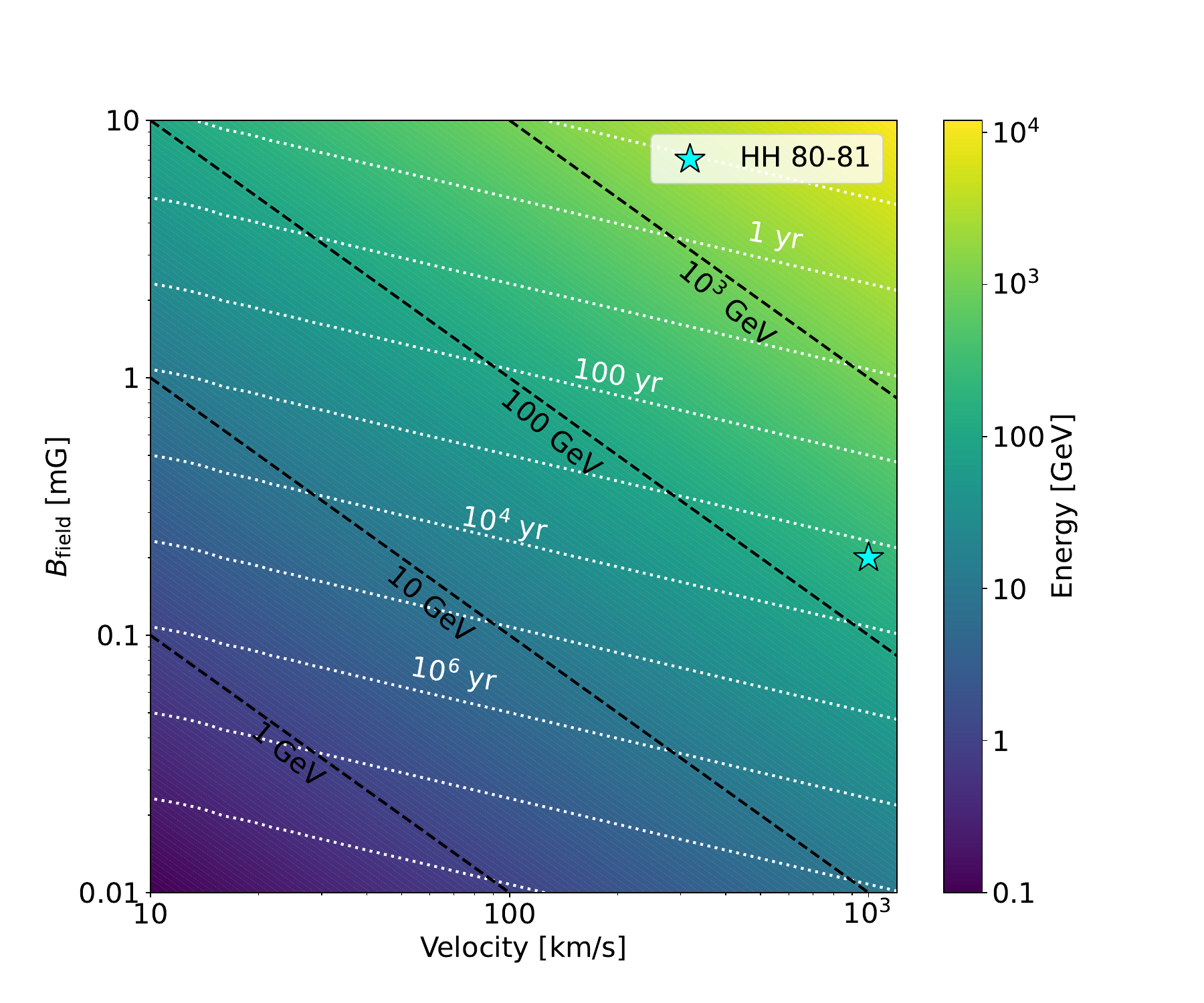}
    \caption{{\bf Maximum energy achieved using the Hillas criterion}. The dashed black lines mark the lines of equal maximum energy as depicted in the color scale. The dotted white lines show the synchrotron cooling times for each of the energies. The cyan star marks the maximum energy that could be achieved by HH 80-81.}
    \label{fig:hillas_maximum_energy}
\end{figure}

Even though this result is dependent on the radius of the region, we can see that for the test case of HH 80-81, in which we have some estimates of the magnetic field \cite{RodriguezKamenetzky2025s} and shock speed of the jet \cite{Marti1993s, 2024Ballys}, we can reach the needed energy to accelerate particles producing the $\gamma$-ray emission. Likewise, for other sources that need to achieve tens-to-hundreds of GeV (or even beyond TeV) energies to produce the observed spectra, the only possibility remains that shocks have very-high speeds ($>100$~km/s) or/and are highly magnetized ($>$0.1~mG). The detected $\gamma$-rays require, thus, shock conditions that have strong implications when observing protostellar jets in other wavelengths, i.e., it breaks the degeneracy on observable line ratios \cite{Pudritz2019s}, pointing to high-velocity shocks with strong magnetic fields at least on the GLP population. 
Moreover, electrons injected into such highly magnetized jets would experience significant synchrotron losses, with short cooling timescales (see white lines in Supplementary Fig. \ref{fig:hillas_maximum_energy}), preventing them from reaching the high energies observed. Since our analysis indicates that the underlying radiation mechanism is consistent across the sample (see Fig. \ref{fig:Lgamma_vs_Lbol}), we conclude that only protons can be identified as the primary particle population responsible for the observed $\gamma$ ray emission.

\section{Derivation of the jet mechanical power}
\label{sect:jet_power}

The observed relation between the $\gamma$-ray luminosity and the bolometric thermal luminosity suggests a common acceleration mechanism for all GLPs, resulting in the production of hadronic CRs maintaining a certain efficiency of acceleration ($\epsilon)$, that is, a fixed fraction of the kinetic energy being transferred into particles. Assuming that CR production takes place in the protostellar jets \cite{Padovani2016s}, the kinetic energy distributed in CRs and the posterior radiative $\gamma$-ray cooling will be a certain portion of the jet mechanical power ($L_{\rm jet}$). 

$$L_{\gamma} \propto \epsilon \, \rho \, L_{\rm jet},$$
where $\epsilon \leq 1$.

The exponential index obtained in Eq. (\ref{eq:Lgamma_vs_Lbol}) can be propagated to obtain the relation between the mechanical luminosity of the entire protostellar jet and the thermal radiative luminosity of the protostellar system, resulting in:

$$L_{\rm jet} \propto L_{\rm bol}^{0.79 \pm 0.15}.$$

However, the extrapolation of this result to all types of protostellar jets might be incorrect. Our findings indicate that only a small fraction of protostellar systems are producing $\gamma$ rays, favoring the $\gamma$-ray production and detection in the most energetic systems. Therefore, our sample may be biased to select only the most powerful jets, affecting to our conclusions regarding the jet mechanical power. Future work individually studying the jet mechanical power of these GLPs (e.g. \cite{Fedriani2018s}) will provide key insights into the determination of the acceleration efficient of this objects.

\begin{figure}[h]
    \centering
    \includegraphics[width=0.62\linewidth]{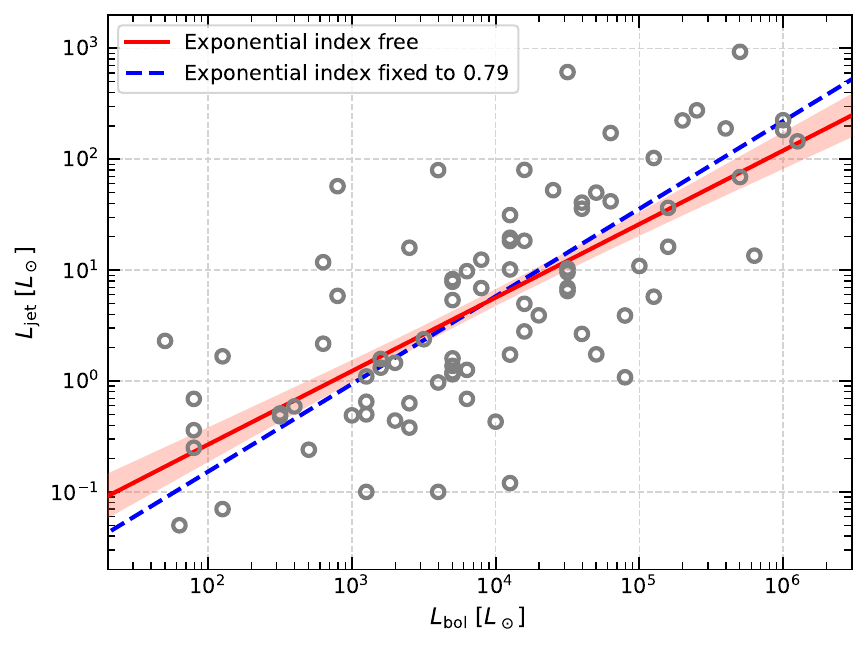}
    \caption{{Relation between $L_{\rm jet}$ and $L_{\rm bol}$}. The red solid line shows the basic linear fitting obtained after freeing the slope and the intercept. Errors from fitting are shown in the red shaded area. Blue dashed line is the resulting fit from fixing the slope to that found in the $L_{\rm bol}$ -- $L_\gamma /\rho$ correlation.}
    \label{fig:Lbol_vs_Ljet}
\end{figure}

Despite this possible bias, we notice that the results obtained through our sample of GLPs are compatible with the results from the APEX Telescope Large Area Survey of the Galaxy in radio wavelengths. Supplementary Fig. \ref{fig:Lbol_vs_Ljet} presents a linear fitting to the data extracted from the COHRS CO (3–2) Survey \cite{2018AtlasGals}, resulting in the following relation:

$$L_{\rm jet} \propto L_{\rm bol}^{0.66 \pm 0.07}.$$

In the same way, an independent survey carried out using $J=3 \xrightarrow{} 2$ of $^{12}$CO and $^{13}$CO with James Clerk Maxwell Telescope \cite{2015Mauds} also studied the relationship between source luminosity and jet mechanical power , resulting in:

$$\frac{L_{\rm jet}}{L_\odot} = 10^{-(2.92 \pm 0.62)} \times \left( \frac{L_{\rm bol}}{L_\odot} \right)^{0.72 \pm 0.15}.$$
 
Obtaining a consistent exponential relationship between $L_{\rm bol}$ and $L_{\rm jet}$ in comparison with an independent methodology is another indicator that protostellar jets can accelerate relativistic particles, superseding thermal emission and shinning through radio and $\gamma$ ray emission --a hallmark of cosmic-ray factories.

\section{Observational biases}
\label{sect:bias}

\begin{figure}[h!]
    \centering
    \includegraphics[width=0.7\linewidth]{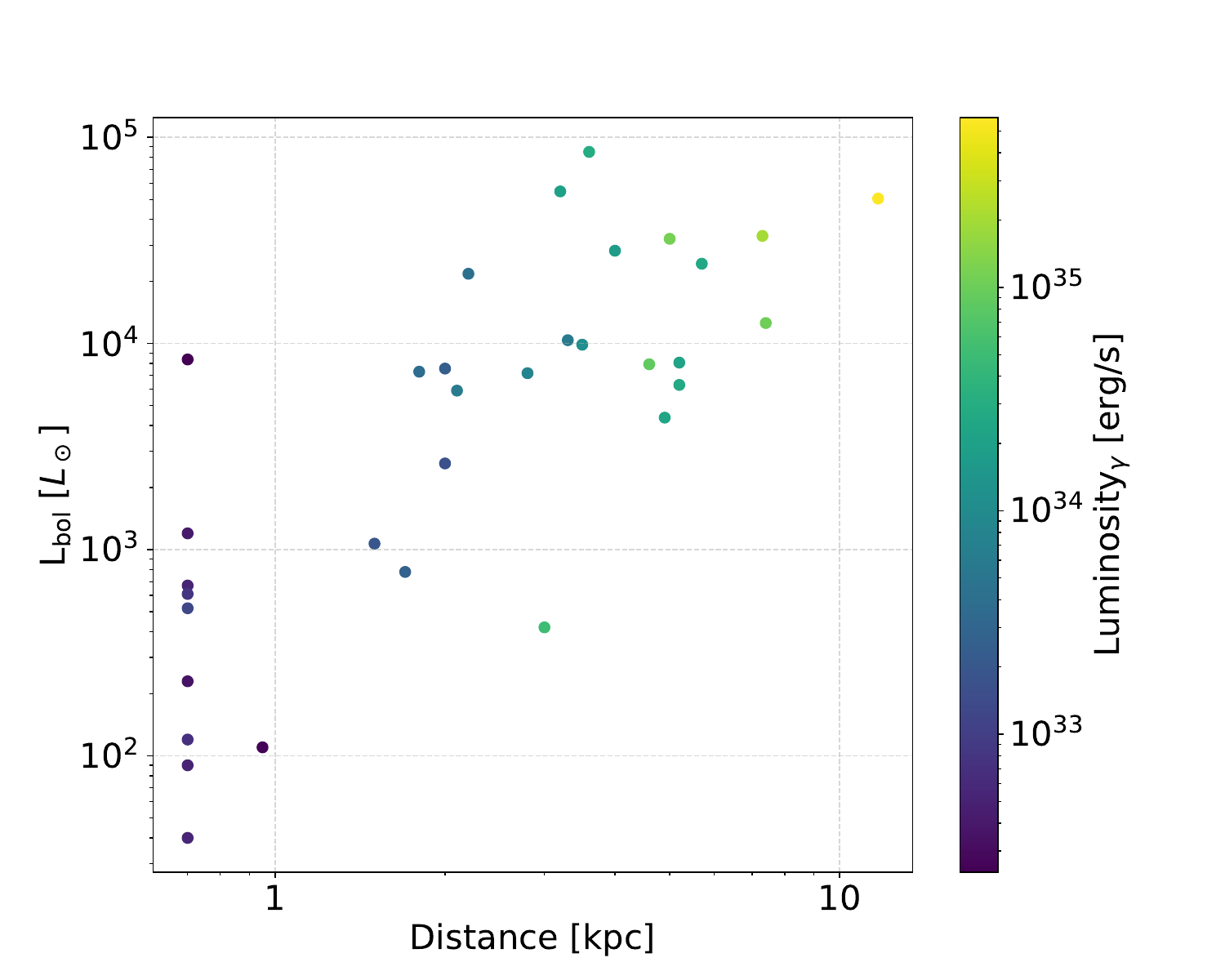}
    \caption{{$L_{\rm bol}$ as a function of the distance to the YSOs}. The color scale represents the $\gamma$-ray luminosity of each associated \textit{Fermi} source.}
    \label{fig:luminosity_vs_distance}
\end{figure}

One of the main results shown in the paper is to establish the relation between the $\gamma$-ray luminosity (originating in accelerated CRs iterating with the ambient material) and the bolometric luminosity of the YSO system (mainly produced by accretion and gravitational collapse), shown on Fig. \ref{fig:Lgamma_vs_Lbol}.
 We also studied the correlation between $L_{\rm bol}$ and the distance for all the GLPs. Supplementary Fig. \ref{fig:luminosity_vs_distance} shows this relation, together with the $\gamma$-ray luminosity as a color scale. We can see that, apart from the nearest sources (located all of them at 0.7 kpc), there is a slight correlation between the bolometric luminosity and the distance, implying that closeby sources are detected with low or high bolometric luminosities, but the further these sources are located, the brighter they need to be for us to detect them in $\gamma$ rays.

\begin{figure}[h!]
    \centering
    \includegraphics[width=0.65\linewidth]{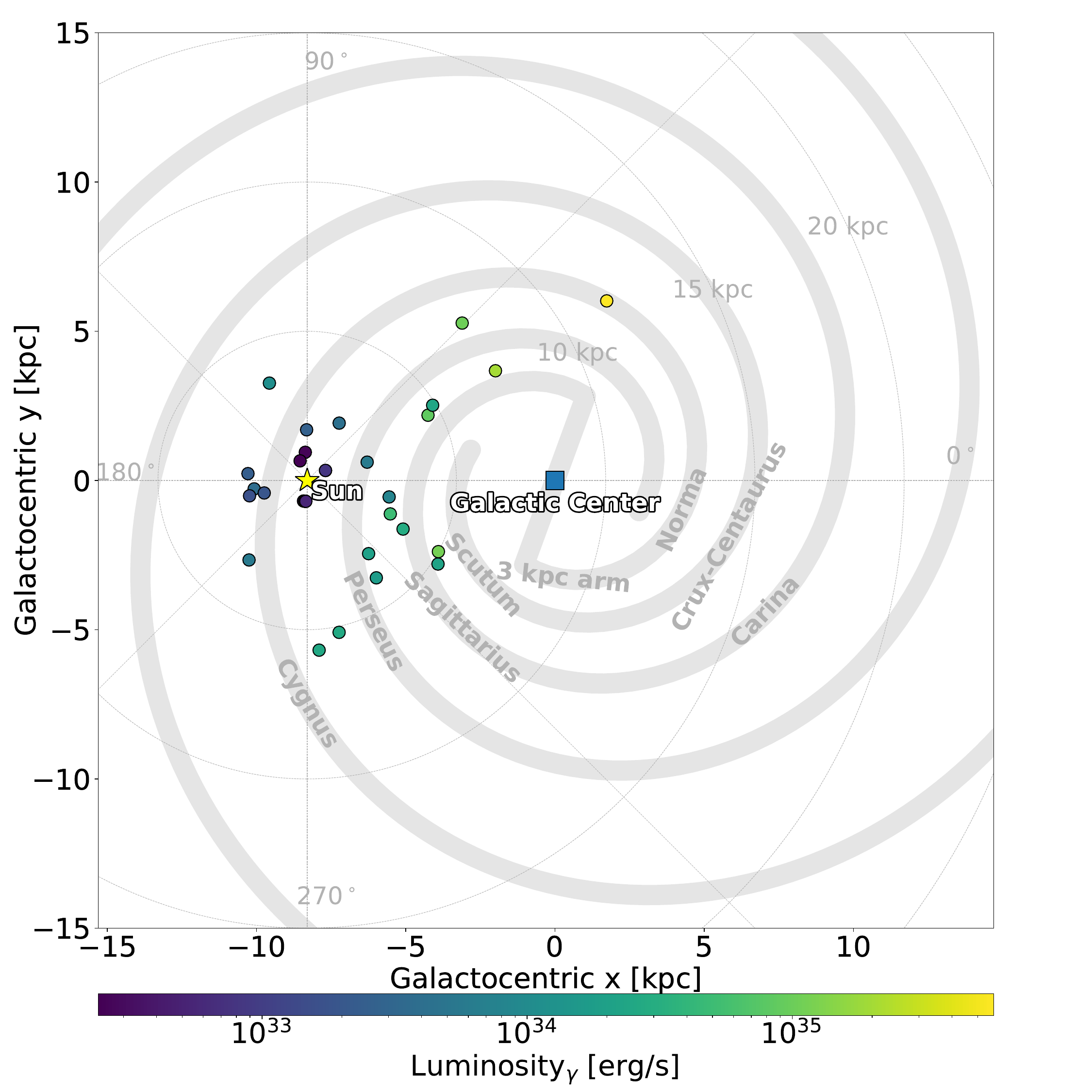}
    \caption{GLP distribution in Galactocentric coordinates sketched with the spiral arm distribution of the Galaxy. The blue square and the yellow star indicate the position of the Galactic Center and the Sun, respectively. The circles color scale represents the $\gamma$-ray luminosity of each associated \textit{Fermi} source.  }
    \label{fig:galactocentric_sketch}
\end{figure}

This is more clearly seen on Supplementary Fig. \ref{fig:galactocentric_sketch}, in which we show a sketch of the Milky Way and its spiral arms, including all GLPs, this time with the $\gamma$-ray luminosity shown as a color scale. We can see that all GLPs are either close or very luminous ones, implying that there is an observation bias in the associated sample. Even though we do not expect all YSOs to be GLPs, the above-mentioned results point towards a non-complete sample of GLPs limited by the sensitivity of current $\gamma$-ray instruments.

\section{Multiwavelength analysis on associated targets}
\label{sect:assoc+}

We performed a multiwavelength analysis of the associated sources to strengthen the conclusions drawn in this work. 

\subsection{Radio surveys}

Protostellar jets are often associated with thermal radio structures, although many of these sources also exhibit non-thermal signatures, such as a negative radio spectral index \cite{Obonyo2024s}. Currently, there is clear evidence in the radio-astronomy literature for non-thermal emission in protostellar jets. The clearest example might be HH\, 80--81, where linearly polarized radiation arises from synchrotron emission of ultrarelativistic electrons \cite{CarrascoGonzalez2010s}. Other studies analyze the spectral index of radio jets to identify synchrotron signatures (e.g., W3(OH) \cite{1999Wilners},  IRAS 16547-4247 \cite{2003Garays}, and HOPS370 \cite{2017Osorios}). Collectively, these works highlight the historical importance of radio astronomy in uncovering non-thermal emission from YSOs.

The NRAO VLA Sky Survey (NVSS) \cite{1998NVSSs} is a comprehensive radio catalog containing more than one million point-like sources, covering the sky north of $-40^\circ$ declination. Although the extensive coverage facilitates cross-matching with RMS sources (see Supplementary Fig. \ref{fig:coverage}), the rms noise level of the Stokes I parameter in NVSS is $\sim$0.45 mJy/beam, which is insufficient for detecting protostellar jets \cite{Anglada2018s}. We detect only the protostellar radio cores of IRAS 07280-1829 and IRAS 22543+6145 (Cepheus A HW2, a very studied radio jet \cite{1994Rodriguezs}) among the 33 GLP sample that we have associated. 

\begin{figure}[h]
    \centering
    \includegraphics[width=0.8\linewidth]{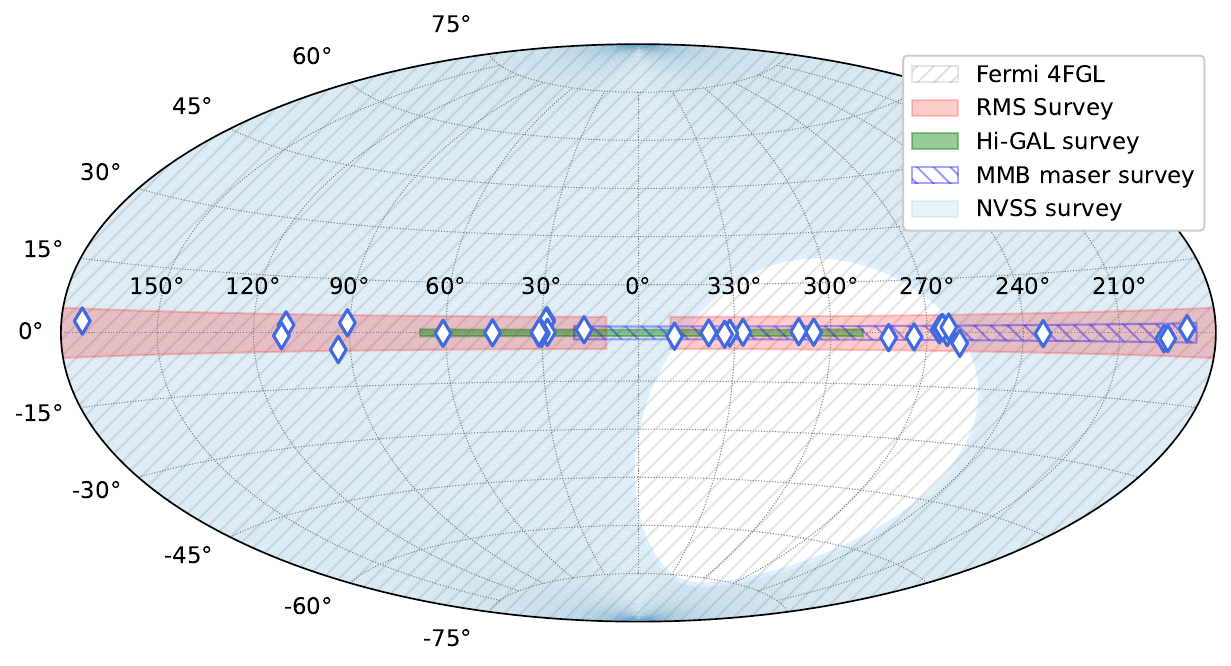}
    \caption{{Coverage of the different surveys used in this work}. Associated GLPs are marked with a blue diamond. Celestial sphere is represented in Galactic coordinates.}
    \label{fig:coverage}
\end{figure}

To complement our radio analysis, we cross-matched our 33 GLPs with the 49 RMS sources observed with ATCA \cite{2016Pursers}, finding five protostellar objects in common. All of them exhibit radio emission. Clear jet structures were reported in IRAS 08513-4201, IRAS 08470-4321, and IRAS 13481-6124. G265.1438+01.4548 shows signatures of being either a protostellar jet or a disk-wind structure, while IRAS 09230-5148 was initially detected as a highly compact HII region, although more recent studies have reported evidence of a protostellar jet \cite{2022Kumars}. 

\subsection{Multiwavelength studies of individual sources}

In addition to the extended survey performed with ATCA, we also found jet-like signatures using CO observations in IRAS 06103+1523 \cite{2006Kims}, IRAS 07280-1829 \cite{2008Gyulbudahgians}, IRAS 08588-4347 \cite{1999Yamaguchis}, IRAS 18274-0212 \cite{2014Dunhams}, and IRAS 21454+4718 \cite{1983Levreaults}.
Additionally, complex analyses on IRAS 16172-5028 with multiple molecule lines including SiO suggest the presence of an outflow driven by a very early stage protostar \cite{2007Los}. G030.1981-00.169 protostellar jet was detected using formaldehyde molecular lines \cite{2024Ortegas}. In the same way, interferometric radio continuum with VLA observations have reported a protostellar jet driven by IRAS 23139+6145 \cite{2006Trinidads}.

Complementing these results with near and far infrared high-sensitivity and resolution techniques, jet structures have been detected in the cases of IRAS 06058+2138 \cite{2024Crowes}, IRAS 08211-4158 (HH 219) \cite{2004Carattis}, IRAS 21007+5036 \cite{2013Walawenders}, and G173.4815+02.4459 \cite{2015Navaretes}. The emission near-infrared lines of Fe II and H$_2$ points towards the potential presence of a protostellar jet, such as in the cases of G045.4641+00.0284 and G173.4815+02.4459 \cite{2013Coopers}.

\subsection{Hi-GAL survey}

In the infrared range, the Herschel infrared Galactic Plane Survey (Hi-GAL) presents a more comprehensive catalog of protostars and pre-stellar objects when compared with the MSX catalog used in this work, which results from more sensitive observations in the mid and far infrared wavelength \cite{2010HiGALs, 2017Elias}. It compiles more than 60,000 sources, $\sim$25,000 are classified as protostars (and potentially protostellar jets). Unfortunately, the spatial coverage of Hi-GAL is more limited than that of the RMS survey (see Supplementary Fig. \ref{fig:coverage}), which constrains a direct comparison with the GLP population.

\begin{figure}[h!]
    \centering
    \includegraphics[width=0.75\linewidth]{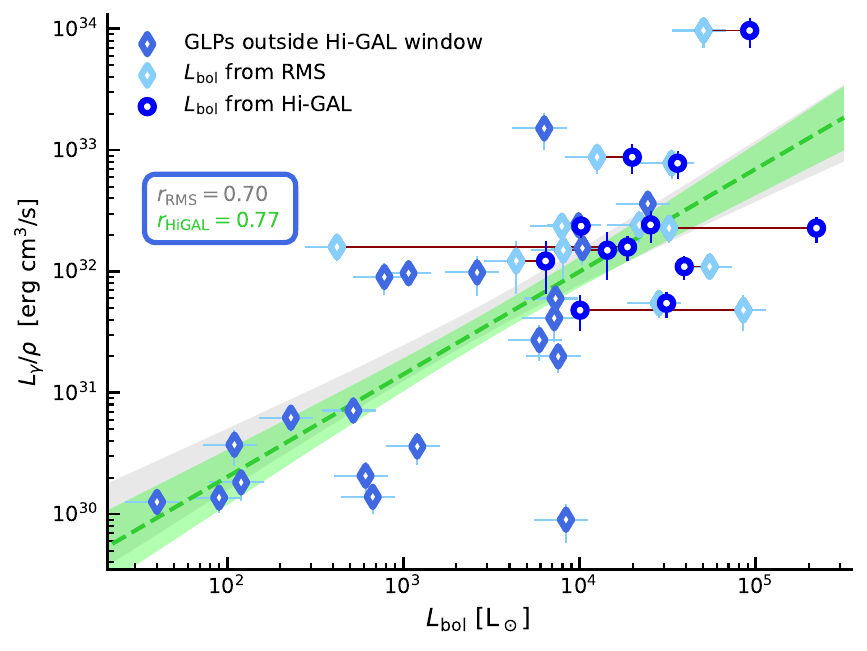}
    \caption{{$L_{\rm bol}$ -- $L_\gamma /\rho$ correlation with corrected $L_{\rm bol}$ from Hi-GAL}. Pearson's coefficients are indicated in legend. The green dashed line show the updated correlation of the combination of Hi-GAL+RMS data (uncertainties represented by green shaded area). In contrast, grey shaded region show the correlation using only RMS luminosities. Dark-red horizontal lines relate the old RMS luminosities (diamonds) with those compiled in Hi-GAL (circles), and their corresponding errors extracted from both catalogs (light-blue errorbars).}
    \label{fig:higal_corrections}
\end{figure}

The average $\rho_{\rm cpt}$ derived from Hi-GAL YSOs is $\sim 2.5\times 10^3$ sources deg$^{-2}$, approximately thirty times higher than that found for the RMS survey over the same observational window. This, combined with the small spatial window and the correspondingly reduced number of 4FGL sources, complicates the association analysis of Hi-GAL YSOs unless highly restrictive cuts are applied in the derived protostellar parameters. Nevertheless, and as a sanity check, we cross-matched the RMS and Hi-GAL surveys and obtained the same 12 associations as we did with the RMS data under the Hi-GAL observational window.

Hi-GAL provides more spectral flux points and combines far and mid-infrared bands, allowing for improved bolometric luminosity estimates. Supplementary Fig.~\ref{fig:higal_corrections} represents the corrections obtained when adopting the updated $L_{\rm bol}$. We recalibrate the new $L_{\rm bol}$ using RMS distance values, since some of these are based on maser parallaxes, which are more accurate than the kinematic distances used by Hi-GAL. One may notice that most of the original RMS luminosities and those from Hi-GAL are quite similar; then, only a small number of sources display visible connection lines between the old and new data points. However, applying these corrections improves the correlations in the $L_{\gamma}$–$L_{\rm bol}$ relation and reduces the number of outliers, with 11 out of the 12 Hi-GAL sources showing compatible luminosities or moving closer to the relation given by Eq. (\ref{eq:Lgamma_vs_Lbol}).

Supplementary Fig. \ref{fig:higal_corrections} combines the associated GLPs outside the spatial region of Hi-GAL plus the 12 corrected sources. While we obtain a consistent $L_\gamma - L_{\rm bol}$ relation, the Pearson's coefficient improves from $\sim$0.70 to $\sim$0.77.

\subsection{MMB maser survey}

\begin{figure}[h]
    \centering
    \includegraphics[width=0.75\linewidth]{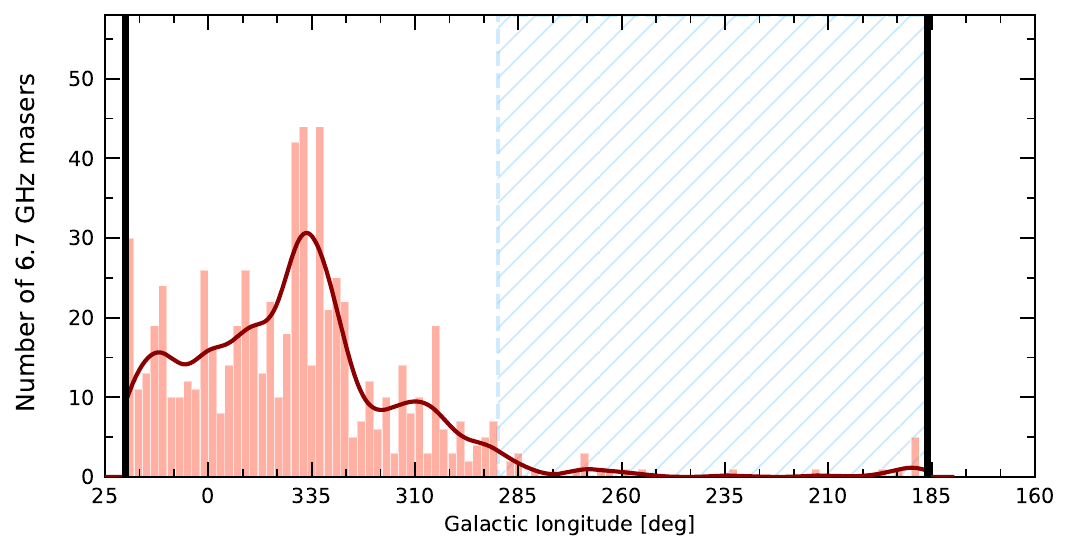}
    \caption{{Methanol maser distribution in the observational window of the MMB survey}. Hatched area shows the spatial region with almost null abundance of masers. Thick-black lines show the observational range of the survey. Red solid line show the smoothed distribution of masers.}
    \label{fig:maser_distrib}
\end{figure}

It is well established in the current literature that the accretion activity of massive star-forming objects powers 6.7 GHz methanol masers (see e.g. \cite{2017Moscadellis, 2018Hunters, 2018Szymczaks, 2023Burnss}). We therefore expect to find methanol masers in some of our associated GLPs, tracing massive YSOs with important accretion activity. We made use of the Methanol Multibeam (MMB;~\cite{2010Caswells, 2010Greens, 2011Caswells, 2012Greens}) survey to identify which GLPs have 6.7 GHz methanol masers associated. We found irregularities in the spatial distribution of maser sources, indicating possible biases with the coverage or sensitivity of the MMB catalog. For instance, Supplementary Fig. \ref{fig:maser_distrib} illustrate the lack of sources within the covered range of $186^\circ \lesssim  \ell < 290^\circ$. We thus decided to focus in the spatial window of $\ell \in [0^\circ, 20^\circ) \cup (290^\circ, 360^\circ)$, that contains 8 potential GLPs. We obtain that six out of the eight GLPs have at least one maser inside the 95\% confidence area.

\section{Investigation of multiwavelength counterparts}
\label{sect:mw}

To prove that our sample of associated YSOs is the most likely origin of the detected $\gamma$-ray sources, we perform a systematic search of other potential emitters in the surroundings of our selected 33 GLPs. We are considering as potential $\gamma$-ray sources all the supernova remnants compiled in the Green's Galactic supernova remnant catalog \cite{2025Greens}; the pulsars detected in the Australia Telescope National Facility (ATNF) pulsar catalog \cite{2005ATNFs}; and the pulsar wind nebula recorded in the Pulsar Wind Nebula catalog \cite{2006PWNcats}. We also consider the young stellar  open clusters ($<$30 Myr) catalog compiled using data from Gaia DR2 \cite{2024Cellis}. In the case of pulsars, we restrict their potential to emit $\gamma$ rays to those with an energy-loss rate exceeding that of the associated $\gamma$-ray source. A similar criterion is applied to young stellar clusters, for which we exclude all systems whose combined wind luminosity is lower than the luminosity of the $\gamma$-ray source. Additionally, since all our 33 GLPs are located close to the Galactic plane where extinction can hide extragalactic objects, we employ the Radio Fundamental Catalog \cite{2025RFCs} for compact and bright extragalactic sources that may indicate the presence of an active galactic nuclei in our regions of interest. 

\begin{figure}[p]
    \centering
    \includegraphics[width=0.65\linewidth]{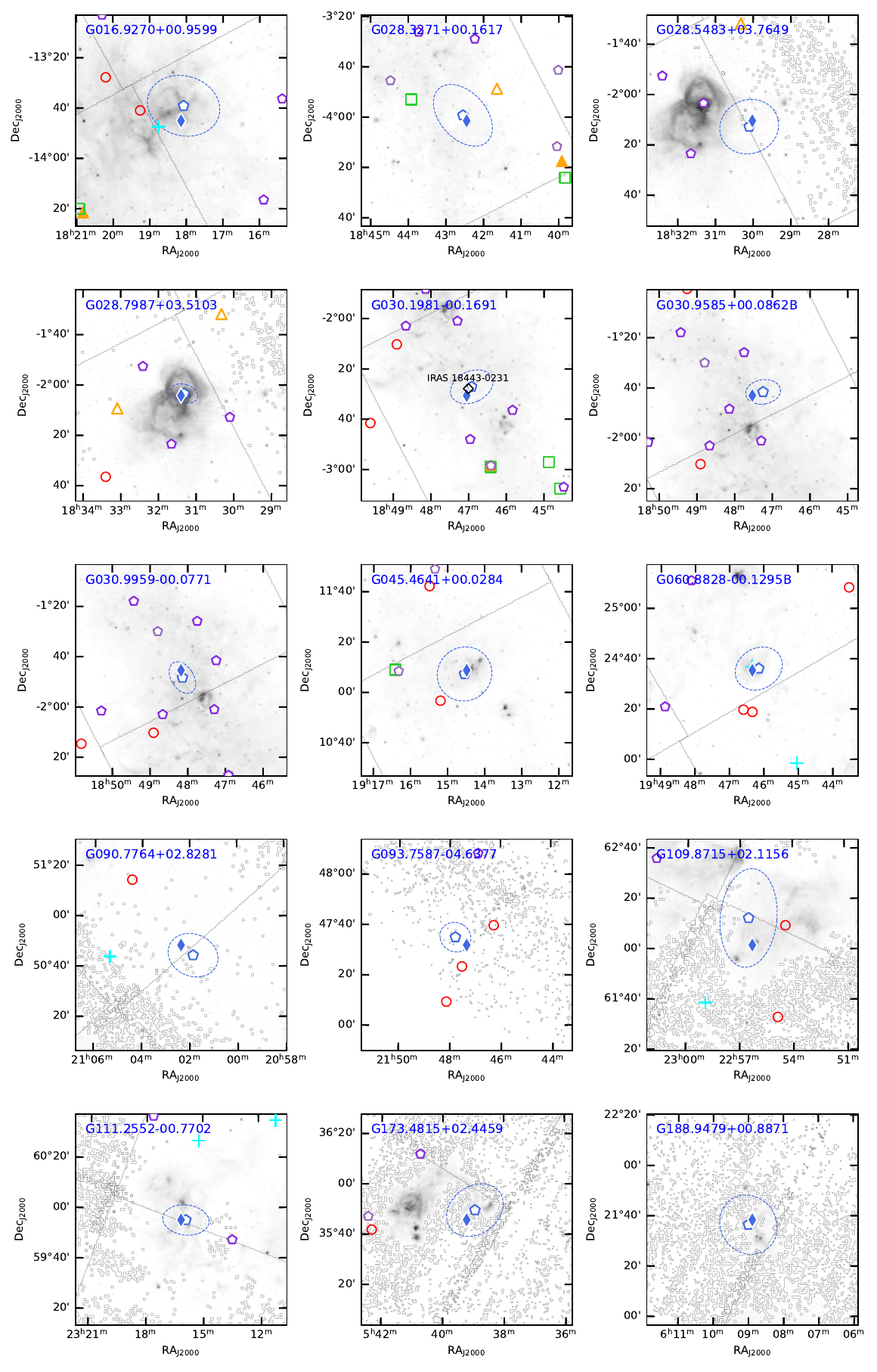}
    \caption{
    Positional analysis of 15 selected 4FGL sources and nearby potential $\gamma$-ray emitters. Blue pentagons (with 99\% ellipses) mark 4FGL associations and blue filled diamonds indicate YSOs (GLP candidates). Yellow triangles denote pulsars (pulsar wind nebulae are filled), red circles compact radio sources, green squares supernova remnants, and cyan crosses young stellar clusters. Dark- (light-) purple pentagons show unassociated (associated) 4FGL sources. Background: MSX $8,\mu$m maps.}
    \label{fig:mw1}
\end{figure}

\begin{figure}[p]
    %\ContinuedFloat
    \centering
    \includegraphics[width=0.65\linewidth]{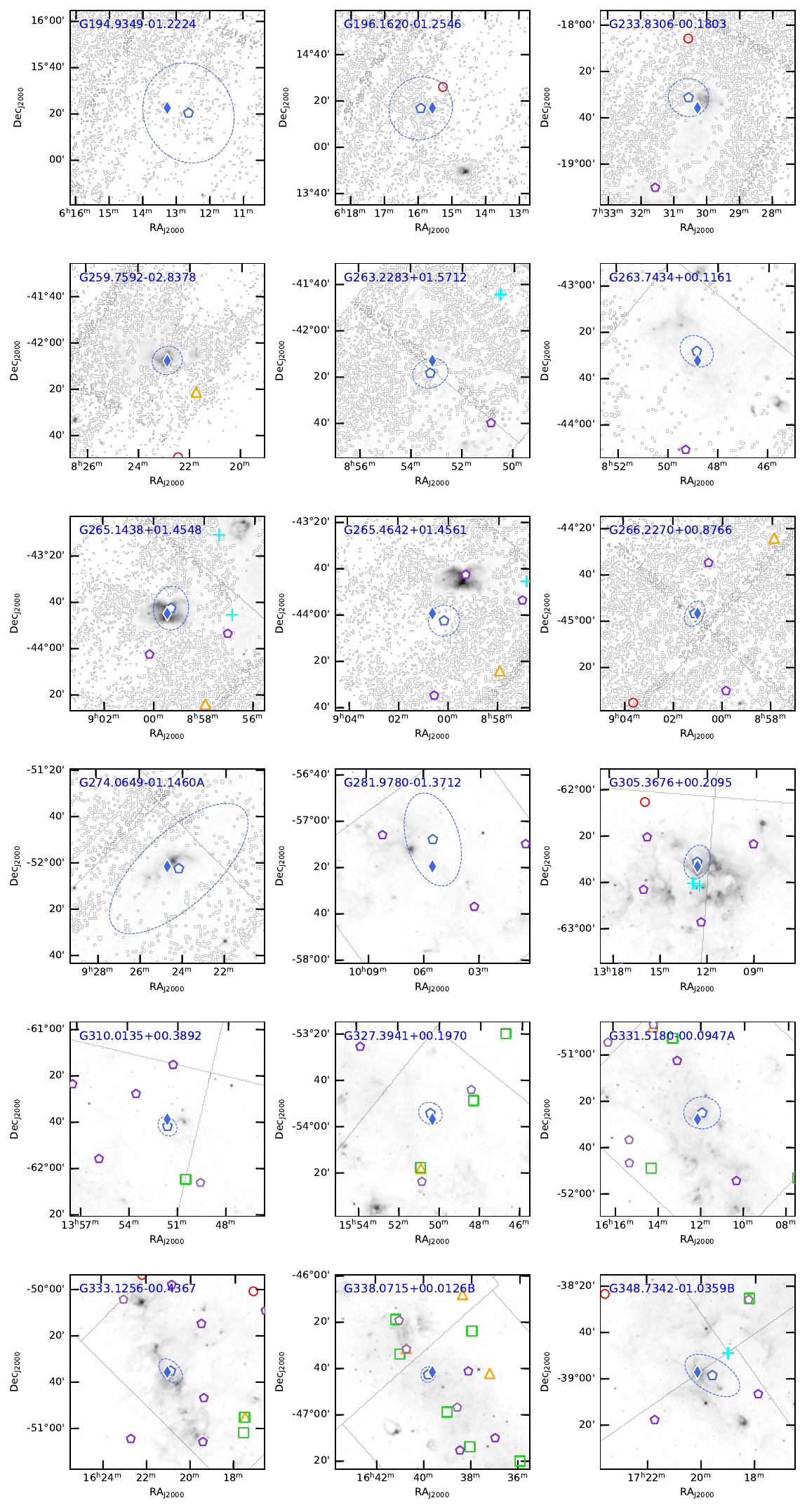}
    \caption{Positional analysis of the remaining 18 associations and nearby potential $\gamma$-ray emitters. Blue pentagons (with 99\% ellipses) mark 4FGL associations and blue filled diamonds indicate YSOs (GLP candidates). Yellow triangles denote pulsars (pulsar wind nebulae are filled), red circles compact radio sources, green squares supernova remnants, and cyan crosses young stellar clusters. Dark- (light-) purple pentagons show unassociated (associated) 4FGL sources. Background: MSX $8,\mu$m maps.}
    \label{fig:mw2}
\end{figure}

Supplementary Figs. \ref{fig:mw1} \& \ref{fig:mw2} show that in only three out of the 33 selected GLPs, we see another counterpart within the 99\% confidence area for the position of the 4FGL sources. It is worth noting that this value is consistent with the expected number of false detections in our sample.

%%========================%%
%%     Supp Reference     %%
%%========================%%
\clearpage
\renewcommand{\refname}{Supplementary references}

\end{document}